\documentclass[usenatbib]{mn2e}
\usepackage{amssymb,amsmath,graphicx,natbib,setspace,pifont}
\newcommand{\HI}{\rm{HI}}
\newcommand{\HII}{\rm{HII}}
\newcommand{\Hm}{\rm{H_{2}} }
\newcommand{\NHI}{{N_{\rm HI}}}
\newcommand{\fNHI}{f(N_{\rm HI},z)}

\newcommand{\ps}{\,{\rm s^{-1}}}
\newcommand{\cmsq}{\,{\rm cm^{-2}}}
\newcommand{\cms}{\,{\rm cm^{2}}}
\newcommand{\cmc}{\,{\rm cm^{3}}}
\newcommand{\cmcb}{\,{\rm cm^{-3}}}

\newcommand{\nH}{n_{\rm _{H}} }

\newcommand{\nHI}{n_{\rm _{HI}}}
\newcommand{\nHII}{n_{\rm _{HII}} }

\newcommand{\nel}{n_{\rm e} }

\newcommand{\Ob}{\Omega_{\rm b} }
\newcommand{\Om}{\Omega_{\rm m} }

\newcommand{\Ol}{\Omega_{\Lambda} }
\newcommand{\ns}{n_{\rm s} }
\newcommand{\sigeight}{\sigma_{\rm 8} }

\newcommand{\Msun}{{\rm M_{\odot}} }
\newcommand{\Msunh}{h^{-1} {\rm M_{\odot}} }
\newcommand{\Mpch}{h^{-1} {\rm Mpc} }
\newcommand{\kpch}{h^{-1} {\rm kpc} }

\newcommand{\Gadget}{{\small GADGET-3} }

\newcommand{\TRAPHIC}{{\small TRAPHIC}~}

\newcommand{\apjl}{{ApJL}}
\newcommand{\apj}{{ApJ}}

\newcommand{\mnras}{{MNRAS}}
\newcommand{\aap}{{A\&A}}

\newcommand{\cmark}{\ding{51}}
\newcommand{\xmark}{\ding{55}}

\begin{document}
\title[On the evolution of the HI CDDF]{On the evolution of the HI column density distribution in cosmological simulations}
\author[A.~Rahmati et al.]
  {Alireza~Rahmati$^1$\thanks{rahmati@strw.leidenuniv.nl}, Andreas H. Pawlik$^2$, Milan Rai\v{c}evi\`{c}$^1$, Joop Schaye$^1$\\
  $^1$Leiden Observatory, Leiden University, P.O. Box 9513, 2300 RA Leiden, The Netherlands\\
  $^2$Max-Planck Institute for Astrophysics, Karl-Schwarzschild-Strasse 1, 85748 Garching, Germany}

\maketitle

\begin{abstract}
We use a set of cosmological simulations combined with radiative transfer calculations to investigate the distribution of neutral hydrogen in the post-reionization Universe. We assess the contributions from the metagalactic ionizing background, collisional ionization and diffuse recombination radiation to the total ionization rate at redshifts $z=0-5$. We find that the densities above which hydrogen self-shielding becomes important are consistent with analytic calculations and previous work. However, because of diffuse recombination radiation, whose intensity peaks at the same density, the transition between highly ionized and self-shielded regions is smoother than what is usually assumed. We provide fitting functions to the simulated photoionization rate as a function of density and show that post-processing simulations with the fitted rates yields results that are in excellent agreement with the original radiative transfer calculations. The predicted neutral hydrogen column density distributions agree very well with the observations. In particular, the simulations reproduce the remarkable lack of evolution in the column density distribution of Lyman limit and weak damped Ly$\alpha$ systems below z = 3. The evolution of the low column density end is affected by the increasing importance of collisional ionization with decreasing redshift. On the other hand, the simulations predict the abundance of strong damped Ly$\alpha$ systems to broadly track the cosmic star formation rate density.
\end{abstract}


\section{Introduction}

A substantial fraction of the interstellar medium (ISM) in galaxies consists of atomic hydrogen. This makes studying the distribution of neutral hydrogen (HI) and its evolution crucial for our understanding of various aspects of star formation. In the local universe, the HI content of galaxies is measured through 21-cm observations, but at higher redshifts this will not be possible until the advent of significantly more powerful telescopes such as the Square Kilometer Array\footnote{http://www.skatelescope.org/}. However, at $z \lesssim 6$, i.e., after reionization, the neutral gas can already be probed through the absorption signatures imprinted by the intervening HI systems on the spectra of bright background sources, such as quasars. 
\par
The early observational constraints on the HI column density distribution function (HI CDDF hereafter), from  quasar absorption spectroscopy at $z \lesssim 3$, were well described by a single power-law in the range $\NHI \sim 10^{13}-10^{21} \cmsq$ \citep{Tytler87}. Thanks to a significant increase in the number of observed quasars and improved observational techniques, more recent studies have extended these observations to both lower and higher HI column densities and to higher redshifts \citep[e.g.,][]{Kim02, Peroux05, Omeara07,Noterdaeme09, Prochaska09, PW09, Omeara12,Noterdaeme12}. These studies have revealed a much more complex shape which has been described using several different power-law functions \citep[e.g.,][]{Prochaska10, Omeara12}. 
\par
The shape of the HI CDDF is determined by both the distribution and ionization state of hydrogen. Consequently, determining the distribution function of HI column densities  requires not only accurate modeling of the cosmological distribution of gas, but also radiative transfer (RT) of ionizing photons. As a starting point, the HI CDDF can be modeled by assuming a certain gas profile and exposing it to an ambient ionizing radiation field \citep[e.g.,][]{Petitjean92, Zheng02}. Although this approach captures the effect of self-shielding, it cannot be used to calculate the detailed shape and normalization of the HI CDDF which results from the cumulative effect of large numbers of objects with different profiles, total gas contents, temperatures and sizes. Moreover, the interaction between galaxies and the circum-galactic medium through accretion and various feedback mechanisms, and its impact on the overall gas distribution are not easily captured by simplified models. Therefore, it is important to complement these models with cosmological simulations that model the evolution of the large-scale structure of the Universe and the formation of galaxies.

The complexity of the RT calculation depends on the HI column density. At low HI column densities (i.e., $\NHI \lesssim 10^{17} \cmsq$, corresponding to the so-called Lyman-$\alpha$ forest), hydrogen is highly ionized by the metagalactic ultra-violet background radiation (hereafter UVB) and largely transparent to the ionizing radiation. For these systems, the HI column densities can therefore be accurately computed in the optically thin limit. At higher HI column densities (i.e., $\NHI \gtrsim 10^{17} \cmsq$, corresponding to the so-called Lyman Limit and Damped Lyman-$\alpha$ systems), the gas becomes optically thick and self-shielded. As a result, the accurate computation of the HI column densities in these systems requires precise RT simulations. On the other hand, at the highest $\HI$ column densities where the gas is fully self-shielded and the recombination rate is high, non-local RT effects are not very important and the gas remains largely neutral. At these column densities, the hydrogen ionization rate may, however, be strongly affected by the local sources of ionization (\citealp{Miralda05,Schaye06}; Rahmati et al. in prep.). In addition, other processes like $\rm{H}_2$ formation \citep{Schaye01b, Krumholz09,Altay11} or mechanical feedback from young stars and / or AGNs \citep{Erkal12}, can also affect the highest HI column densities.
\par
Despite the importance of RT effects, most of the previous theoretical works on the HI column density distribution did not attempt to model RT effects in detail \citep[e.g.,][]{Katz96, Gardner97, Haehnelt98,Gardner01,Cen03,Nagamine04,Nagamine07}. Only very recent works incorporated RT, primarily to account for the attenuation of the UVB \citep{Razoumov06, Pontzen08, Fumangalli11,Altay11,McQuinn11} and found a sharp transition between optically thin and self-shielded gas that is expected from the exponential nature of extinction. 
\par
The aforementioned studies focused mainly on redshifts $z = 2-3$, for which observational constraints are strongest, without investigating the evolution of the HI distribution. They found that the HI CDDF in current cosmological simulations is in reasonable agreement with observations in a large range of HI column densities. Only at the highest HI column densities (i.e., $\NHI \gtrsim 10^{21} \cmsq$) the agreement is poor. However, it is worth noting that the interpretation of these HI systems is complicated due to the complex physics of the ISM and ionization by local sources. Moreover, the observational uncertainties are also larger for these rare high $\NHI$ systems.
\par
In this paper, we investigate the cosmological HI distribution and its evolution during the last $\gtrsim 12$ billion years (i.e., $z \lesssim 5$). For this purpose, we use a set of cosmological simulations which include star formation, feedback and metal-line cooling in the presence of the UVB. These simulations are based on the Overwhelmingly Large Simulations (OWLS) presented in \citet{Schaye10}. To obtain the HI CDDF, we post-processed the simulations with RT, accounting for both ionizing UVB radiation and ionizing recombination radiation (RR). In contrast to previous works, we account for the impact of recombination radiation explicitly, by propagating RR photons. Using these simulations we study the evolution of the HI CDDF in the range of redshifts $z = 0 - 5$ for column densities $\NHI \gtrsim 10^{16} \cmsq$. We discuss how the individual contributions from the UVB, RR and collisional ionization to the total ionization rate shape the HI CDDF and assess their relative importance at different redshifts.
\par
The structure of this paper is as follows. In \S\ref{sec:ingredients} we describe the details of the hydrodynamical simulations and of the RT, including the treatment of the UVB and recombination radiation. In \S\ref{sec:results} we present the simulated HI CDDF and its evolution and compare it with observations. In the same section we also discuss the contributions of different ionizing processes to the total ionization rate and provide fitting functions for the total photoionization rate as a function of density which reproduce the RT results. Finally, we conclude in \S\ref{sec:conclusions}.

\section{Simulation techniques}
\label{sec:ingredients}
\subsection{Hydrodynamical simulations}
\label{sec:hydro}
We use density fields from a set of cosmological simulations performed using a modified version of the smoothed particle hydrodynamics code \Gadget (last described in \citealp{Springel05}). The subgrid physics is identical to that used in the reference simulation of the OWLS project \citep{Schaye10}. Star formation is pressure dependent and reproduces the observed Kennicutt-Schmidt law \citep{Schaye08}. Chemical evolution is followed using the model of \citet{Wiersma09a}, which traces the abundance evolution of eleven elements by following stellar evolution assuming a \citet{Chabrier03} initial mass function. Moreover, a radiative heating and cooling implementation based on \citet{Wiersma09b} calculates cooling rates element-by-element (i.e., using the above mentioned 11 elements) in the presence of the uniform cosmic microwave background and the UVB model given by \citet{HM01}. About 40 per cent of the available kinetic energy in type II SNe is injected in winds with initial velocity of $600~{\rm{kms^{-1}}}$ and a mass loading parameter $\eta = 2$ \citep{DallaVecchia08}. Our tests show that varying the implementation of the kinetic feedback only changes the HI CDDF in the highest column densities ($\NHI \gtrsim 10^{21} \cmsq$). However, the differences caused by these variations are smaller than the evolution in the HI CDDF and observational uncertainties (see Altay et al. in prep.).
\par
We adopt fiducial cosmological parameters consistent with the most recent WMAP 7-year results: $\Om=0.272,\ \Ob=0.0455,\ \Ol=0.728,\ \sigeight=0.81,\ \ns=0.967$ and $\ h=0.704$ \citep{Komatsu11}. We also use cosmological simulations from the OWLS project which are performed with a cosmology consistent with WMAP 3-year values with $\Om=0.238,\ \Ob=0.0418,\ \Ol=0.762,\ \sigeight=0.74,\ \ns=0.951$ and $\ h=0.73 $. We use those simulations to avoid expensive resimulation with a WMAP 7-year cosmology. Instead, we correct for the difference in the cosmological parameters as explained in Appendix \ref{ap:res-box-cos-tests}.
\par
Our simulations have box sizes in the range $L = 6.25 - 100$ comoving $\Mpch$ and baryonic particle masses in the range $1.7 \times 10^5~\Msunh - 8.7 \times 10^7~\Msunh$. The suite of simulations allows us to study the dependence of our results on the box size and mass resolution. Characteristic parameters of the simulations are summarized in Table \ref{table:hydro_sims}.

\begin{table*}
\caption{List of cosmological simulations used in this work. All the simulations use model ingredients identical to the reference simulation of \citet{Schaye10}. From
  left to right the columns show: simulation identifier; comoving box size;
  number of dark matter particles (there are equally 
  many baryonic particles); initial baryonic particle mass; dark matter
  particle mass; comoving (Plummer-equivalent) gravitational
  softening; maximum physical softening; final redshift; cosmology. The last column shows whether the simulation was post-processed with RT. In simulations without RT, the $\HI$ distribution is obtained by using a fit to the photoionization rates as a function of density measured from simulations with RT.} 
\begin{tabular}{lrcccccccc}
\hline
Simulation & $L$ & $N$ & $m_{\rm b}$ & $m_{\rm dm}$ & $\epsilon_{\rm com}$ & $\epsilon_{\rm prop}$ & $z_{\rm end}$ & Cosmology & RT\\  
& $(\Mpch)$ & & $(\Msunh)$ & $(\Msunh)$ & $(\kpch)$ & &
$(\kpch)$ & &\\
\hline 
\emph{L06N256} &    6.25 & $256^3$ & $1.7 \times 10^5$
& $ 7.9 \times 10^5$ & 0.98 & 0.25 & 2 & WMAP7 & \cmark\\
\emph{L06N128} &    6.25 & $128^3$ & $1.4 \times 10^6$
& $ 6.3 \times 10^6$ & 1.95 & 0.50 & 0 & WMAP7& \cmark\\
\emph{L12N256} &  12.50 & $256^3$ & $1.4 \times 10^6$
& $ 6.3 \times 10^6$ & 1.95 & 0.50 & 2 & WMAP7& \cmark\\
\emph{L25N512} &  25.00 & $512^3$ & $1.4 \times 10^6$
& $ 6.3 \times 10^6$ & 1.95 & 0.50 & 2 & WMAP7& \xmark\\
\emph{L06N128-W3} &  6.25 & $128^3$ & $1.4 \times 10^6$
& $ 6.3 \times 10^6$ & 1.95 & 0.50 & 2 & WMAP3& \cmark\\
\emph{L25N512-W3} &  25.00 & $512^3$ & $1.4 \times 10^6$
& $ 6.3 \times 10^6$ & 1.95 & 0.50 & 2 & WMAP3& \xmark\\
\emph{L25N128-W3} &  25.00 & $128^3$ & $8.7 \times 10^7$
& $ 4.1 \times 10^8$ & 7.81 & 2.00 & 0 & WMAP3& \cmark\\
\emph{L50N256-W3} &  50.00 & $256^3$ & $8.7 \times 10^7$
& $ 4.1 \times 10^8$ & 7.81 & 2.00 & 0 & WMAP3& \cmark\\
\emph{L50N512-W3} &  50.00 & $512^3$ & $1.1 \times 10^7$
& $ 5.1 \times 10^7$ & 3.91 & 1.00 & 0 & WMAP3& \cmark\\
\emph{L100N512-W3} &  100.00 & $512^3$ & $8.7 \times 10^7$
& $ 4.1 \times 10^8$ & 7.81 & 2.00 & 0 & WMAP3& \xmark\\
\hline
\end{tabular}
\label{table:hydro_sims}
\end{table*}
\subsection{Radiative transfer with \TRAPHIC}
\label{sec:traphic}
The RT is performed using \TRAPHIC \citep{Pawlik08,Pawlik11}. \TRAPHIC is an explicitly photon-conserving RT method designed to transport radiation directly on the irregular distribution of SPH particles using its full dynamic range. Moreover, by tracing photon packets inside a discrete number of cones, the computational cost of the RT becomes independent of the number of radiation sources. \TRAPHIC is therefore particularly well-suited for RT calculation in cosmological density fields with a large dynamical range in densities and large numbers of sources. In the following we briefly describe how \TRAPHIC works. More details, as well as various RT tests, can be found in \citet{Pawlik08,Pawlik11}. 
\par
The photon transport in \TRAPHIC proceeds in two steps: the isotropic emission of photon packets with a characteristic frequency $\nu$ by source particles and their subsequent directed propagation on the irregular distribution of SPH particles. The spatial resolution of the RT is set by the number of neighbors for which we generally use the same number of SPH neighbors used for the underlying hydrodynamical simulations, i.e., $N_{\rm{ngb}} = 48$.  
\par
After source particles emit photon packets isotropically to their neighbors, the photon packets travel along their propagation directions to other neighboring SPH particles which are inside their transmission cones. Transmission cones are regular cones with opening solid angle $ 4\pi/N_{\rm{TC}}$ and are centered on the propagation direction. The parameter $N_{\rm{TC}}$ sets the angular resolution of the RT, and we adopt $N_{\rm{TC}} = 64$. We demonstrate convergence of our results with the angular resolution in Appendix \ref{ap:RT-conv-test}. Note that the transmission cones are defined locally at the transmitting particle, and hence the angular resolution of the RT is independent of the distance from the source.
\par
It can happen that transmission cones do not contain any neighboring SPH particles. In this case, additional particles (virtual particles, ViPs) are placed inside the transmission cones to accomplish the photon transport. The ViPs, which enable the particle-to-particle transport of photons along any direction independent of the spatially inhomogeneous distribution of the particles, do not affect the SPH simulation and are deleted after the photon packets have been transferred.
\par
An important feature of the RT with \TRAPHIC is the merging of photon packets which guarantees the independence of the computational cost from the number of sources. Different photon packets which are received by each SPH particle are binned based on their propagation directions in $N_{\rm{RC}}$ reception cones. Then, photon packets with identical frequencies that fall in the same reception cone are merged into a single photon packet with a new direction set by the weighted sum of the directions of the original photon packets. Consequently, each SPH particle holds at most $N_{\rm{RC}} \times N_{\nu}$ photon packets, where $N_{\nu}$ is the number of frequency bins. We set $N_{\rm{RC}}= 8$ for which our tests yield converged results.
\par
Photon packets are transported along their propagation direction until they reach the distance they are allowed to travel within the RT time step by the finite speed of light, i.e., $c \Delta t$. Photon packets that cross the simulation box boundaries are assumed to be lost from the computational domain. We use a time step $\Delta t = 1~{\rm{Myr}}~\left(\frac{L_{\rm{box}}}{6.25~~\Mpch}\right)~\left(\frac{4}{1+z}\right)~\left(\frac{128}{N_{\rm{SPH}}}\right)$, where $N_{\rm{SPH}}$ is the number of SPH particles in each dimension. We verified that our results are insensitive to the exact value of the RT time step: values that are smaller or larger by a factor of two produce essentially identical results. This is mostly because we evolve the ionization balance on smaller subcycling steps,
and because we iterate for the equilibrium solution, as we discuss below. At the end of each time step the ionization states of the particles are updated based on the number of  absorbed ionizing photons.
\par
The number of ionizing photons that are absorbed during the propagation of a photon packet from one particle to its neighbor is given by $\delta \mathcal{N}_{\rm{abs},\nu} = \delta \mathcal{N}_{\rm{in},\nu}[1- \exp(-\tau(\nu))]$ where $\delta \mathcal{N}_{\rm{in},\nu}$ and $\tau(\nu)$ are, respectively, the initial number of ionizing photons in the photon packet with frequency $\nu$ and the total optical depth of all the absorbing species. In this work we mainly consider hydrogen ionization, but in general the total optical depth is the sum $\tau(\nu) = \sum_{\alpha} \tau_{\alpha}(\nu)$ of the optical depth of each absorbing species (i.e., $\alpha \in \{ \rm{HI, HeI, HeII}\}$). Assuming that neighboring SPH particles have similar densities, we approximate the optical depth of each species using $\tau_{\alpha}(\nu) = \sigma_{\alpha}(\nu)n_{\alpha}d_{\rm{abs}}$, where $n_{\alpha}$ is the number density of species, $d_{\rm{abs}}$ is the absorption distance between the SPH particle and its neighbor and $ \sigma_{\alpha}(\nu)$ is the absorption cross section \citep{Verner96}. Note that ViPs are deleted after each transmission, and hence the photons they absorb need to be distributed among their SPH neighbors. However, in order to decrease the amount of smoothing associated with this redistribution of photons, ViPs are assigned only 5 (instead of 48) SPH neighbors. We demonstrate convergence of our results with the number of ViP neighbors in Appendix~\ref{ap:RT-conv-test}.
\par
At the end of each RT time step, every SPH particle has a total number of ionizing photons that have been absorbed by each species, $\Delta \mathcal{N}_{\rm{abs}, \alpha}(\nu)$. This number is used in order to calculate the photoionization rate of every species for that SPH particle. For instance, the hydrogen photoionization rate is given by:
\begin{equation}
\Gamma_{\rm{HI}} =\frac{ \sum_{\nu}\Delta \mathcal{N}_{\rm{abs},\rm{HI}}(\nu)} {\eta_{\rm{HI}} \mathcal{N}_{\rm{H}}\Delta t },
\label{eq:SPH-gamma} 
\end{equation}
where $\mathcal{N}_{\rm{H}}$ is the total number of hydrogen atoms inside the SPH particle and $\eta_{\rm{HI}} \equiv \nHI / \nH$ is the hydrogen neutral fraction.
\par
Once the photoionization rate is known, the evolution of the ionization states is calculated. For instance, the equation which governs the ionization state of hydrogen is
\begin{equation}
\frac{\eta_{\rm{HI}}}{dt} = \alpha_{\rm{HII}} n_e (1-\eta_{\rm{HI}}) - \eta_{\rm{HI}}(\Gamma_{\rm{HI}} + \Gamma_{e,\rm{H}} n_e),
\label{eq:eta-evolution} 
\end{equation}
where $n_e$ is the free electron number density, $ \Gamma_{e,\rm{H}}$ is the collisional ionization rate and $\alpha_{\rm{HII}}$ is the $\HII$ recombination rate. The differential equations which govern the ionization balance (e.g., equation \ref{eq:eta-evolution}) are solved using a subcycling time step, $\delta t = \min (f\tau_{\rm{eq}}, \Delta t)$ where $\tau_{\rm{eq}} \equiv {\tau_{\rm{ion}}\tau_{\rm{rec}}}/({\tau_{\rm{ion}}+\tau_{\rm{rec}}})$, and $f$ is a dimensionless factor which controls the integration accuracy (we set it to $10^{-3}$), $\tau_{\rm{rec}} \equiv 1/\sum_{i} n_e\alpha_{i}$ and $\tau_{\rm{ion}} \equiv 1/\sum_{i}(\Gamma_{i}+n_e\Gamma_{e,i})$. The subcycling scheme allows the RT time step to be chosen independently of the photoionization and recombination time scales without compromising the accuracy of the ionization state calculations\footnote{Other considerations prevent the use of arbitrarily large RT time steps. The RT assumes that species fractions and hence opacities do not evolve within a RT time step. This approximation becomes increasingly inaccurate with increasing RT time steps. Note that in this work, we iterate for ionization equilibrium which help to render our results robust against changes in the RT time step, as our convergence studies confirm.}.  
\par
We employ separate frequency bins to transport UVB and RR photons. Because the propagation directions of photons in different
frequency bins are merged separately, this allows us to track the individual radiation components, i.e., UVB and RR, and to compute their contributions to the total photoionization rate. The implementation of the UVB and the recombination radiation is described in
$\S$~\ref{sec:UVB-normalization} and $\S$~\ref{sec:recrad} below.
\par
At the start of the RT, the hydrogen is assumed to be neutral.  In addition, we use a common simplification \citep[e.g.,][]{Faucher09, McQuinn10, Altay11} by assuming a hydrogen mass fraction of unity, i.e., we ignore helium (only for the RT). To calculate recombination and collisional ionizations rates, we set, in post-processing, the temperatures of star-forming gas particles with densities $\nH > 0.1\cmcb$ to $T_{\rm{ISM}} = 10^4~\rm{K}$, which is typical of the observed warm-neutral phase of the ISM. This is needed because in our hydrodynamical simulations the star-forming gas particles follow a polytropic equation of state which defines their effective temperatures. These temperatures are only a measure of the imposed pressure and do not represent physical temperatures (see \citealp{Schaye08}). To speed up convergence, the hydrogen at low densities (i.e., $\nH < 10^{-3}~\cmcb$) or high temperatures (i.e., $T > 10^5$ K) is assumed to be in ionization equilibrium with the UVB and the collisional ionization rate (see Appendix \ref{ap:equilib_neut}). Typically, the neutral fraction of the box and the resulting $\HI$ CDDF do not evolve after 2-3 light-crossing times (the light-crossing time for the extended box with $L_{\rm{box}} = 6.25$ comoving $\Mpch$ is $\approx 7.5$ Myr at z = 3).
\begin{table*}
\caption{Hydrogen photoionization rate, absorption cross-section, equivalent gray approximation frequency and the self-shielding density threshold (i.e., based on equation \ref{eq:densitySSH}) for three UVB models: \citet{HM01} (HM01; used in this work), \citet{HM12} (HM12) and \citet{Faucher09} (FG09) at different redshifts. For the calculation of the photoionization rate and absorption cross-sections, only photons with energies below $54.4$ eV are taken into account, effectively assuming that more energetic photons are absorbed by He.}
\begin{tabular}{c c c c c c}
\hline
Redshift & UVB & $\Gamma_{\rm{UVB}}$ (${\rm{s^{-1}}}$)&$\bar{\sigma}_{\nu_{\HI}}$ ($\cms$) & $\mathcal{E}_{\rm{eq}}$ (eV) &$n_{\rm{H,SSh}}$ ($\cmcb$) \\
\hline
\hline
$z =0$ & HM01 & $8.34 \times 10^{-14}$ & $3.27 \times 10^{-18}$ & 16.9 & $1.1 \times 10^{-3}$\\
            & HM12 & $2.27 \times 10^{-14}$ & $2.68 \times 10^{-18}$ & 18.1 & $5.1 \times 10^{-4}$ \\
            & FG09 & $3.99 \times 10^{-14}$ & $2.59 \times 10^{-18}$ & 18.3  & $7.7 \times 10^{-4}$\\
\hline
$z =1$ & HM01 & $7.39 \times 10^{-13}$ & $2.76 \times 10^{-18}$ & 17.9 & $5.1 \times 10^{-3} $\\
            & HM12 & $3.42 \times 10^{-13}$ & $2.62 \times 10^{-18}$ & 18.2 & $3.3 \times 10^{-3}$ \\
            & FG09 & $3.03 \times 10^{-13}$ & $2.37 \times 10^{-18}$ & 18.8 & $3.1 \times 10^{-3}$\\
\hline
$z =2$ & HM01 & $1.50 \times 10^{-12}$ & $2.55 \times 10^{-18}$ & 18.3 & $8.7 \times 10^{-3} $\\
            & HM12 & $8.98 \times 10^{-13}$ & $2.61 \times 10^{-18}$ & 18.2 & $6.1 \times 10^{-3} $\\
            & FG09 & $6.00 \times 10^{-13}$ & $2.27 \times 10^{-18}$ & 19.1 & $5.1 \times 10^{-3} $\\
\hline
$z =3$ & HM01 & $1.16 \times 10^{-12}$ & $2.49 \times 10^{-18}$ & 18.5 & $7.4 \times 10^{-3}$\\
            & HM12 & $8.74 \times 10^{-13}$ & $2.61 \times 10^{-18}$ & 18.2 & $6.0 \times 10^{-3}$\\
            & FG09 & $5.53 \times 10^{-13}$ & $2.15 \times 10^{-18}$ & 19.5 & $5.0 \times 10^{-3} $\\
\hline
$z =4$ & HM01 & $7.92 \times 10^{-13}$ & $2.45 \times 10^{-18}$ & 18.6 & $5.8 \times 10^{-3}$\\
            & HM12 & $ 6.14 \times 10^{-13}$ & $2.60 \times 10^{-18}$ & 18.3 & $4.7 \times 10^{-3} $\\
            & FG09 & $4.31 \times 10^{-13}$ & $2.02 \times 10^{-18}$ & 19.9 & $4.4 \times 10^{-3} $\\
\hline
$z =5$ & HM01 & $5.43 \times 10^{-13}$ & $2.45 \times 10^{-18}$ & 18.6 & $4.5 \times 10^{-3}$\\
            & HM12 & $4.57 \times 10^{-13}$ & $2.58 \times 10^{-18}$ & 18.3 & $3.9 \times 10^{-3} $\\
            & FG09 & $3.52 \times 10^{-13}$ & $1.94 \times 10^{-18}$ & 20.1 & $4.0 \times 10^{-3} $\\
\hline
\end{tabular}
\label{table:UVBz}
\end{table*}

\subsection{Ionizing background radiation}
\label{sec:UVB-normalization}
Although our hydrodynamical simulations are performed using periodic boundary conditions, we use absorbing boundary conditions for the RT. This is necessary because our box size is much smaller than the mean free path of ionizing photons. We simulate the ionizing background radiation as plane-parallel radiation entering the simulation box from its sides. At the beginning of each RT step, we generate a large number of photon packets, $N_{\rm{bg}}$, on the nodes of a regular grid at each side of the simulation box and set their propagation directions perpendicular to the sides. The number of photon packets is chosen to obtain converged results. Furthermore, to avoid numerical artifacts close to the edges of the box, we use the periodicity of our simulations to extend the simulation box by the typical size of the region where we generate the background radiation (i.e., $2\%$ of the box size from each side). These extended regions are excluded from the analysis, thereby removing the artifacts without losing any information contained in the original simulation box.
\par
The photon content of each packet is normalized such that in the absence of any absorption (i.e., assuming the optically thin limit), the total photon density of the box corresponds to the desired uniform hydrogen photoionization rate. If we assume that all the photons with frequencies higher than $\nu_{\rm{HeII}}$ are absorbed by helium, then the hydrogen photoionization rate can be written as:
\begin{eqnarray}
&{}& \Gamma_{\rm{UVB}} ~~= \int_{\nu_{\HI}}^{\nu_{\rm{HeII}}}{4\pi ~\frac{J_{\nu}}{h\nu}~\sigma_{\rm{HI},\nu}~d\nu}  \nonumber \\
&{}& \qquad  \qquad  \equiv \frac{4\pi~\bar{\sigma}_{\nu_{\HI}}}{h}  ~\int_{\nu_{\HI}}^{\nu_{\rm{HeII}}}{\frac{J_{\nu}}{\nu}~d\nu},
\label{eq:Gamma-photon-density}
\end{eqnarray}
where $J_{\nu}$ is the radiation intensity (in units $\rm{erg~cm^{-2}~s^{-1}~sr^{-1}~Hz^{-1}}$), $\nu_{\HI}$ and $\nu_{\rm{HeII}}$ are respectively the frequency at the Lyman-limit and the frequency at the $\rm{HeII}$ ionization edge, and $\sigma_{\rm{HI},\nu}$ is the neutral hydrogen absorption cross-section for ionizing photons. In the last equation we have defined the gray absorption cross-section,
 \begin{equation}
\bar{\sigma}_{\nu_{\HI}} \equiv \frac{ \int_{\nu_{\HI}}^{\nu_{\rm{HeII}}} {J_{\nu}/{\nu}~\sigma_{\rm{HI},\nu}~d\nu} } {\int_{\nu_{\HI}}^{\nu_{\rm{HeII}}}{J_{\nu}/{\nu}~d\nu} }.
\label{eq:gray-sigma}
\end{equation}
\par
The radiation intensity is related to the photon energy density, $u_{\nu}$,
\begin{equation}
J_{\nu} = \frac{u_{\nu}~c}{4\pi} = \frac{n_{\nu}~h\nu~c}{4\pi} ,
\label{eq:Intensity} 
\end{equation}
where $n_{\nu}$ is the number density of photons inside the box. Combining Equations \ref{eq:Gamma-photon-density}-\ref{eq:Intensity} yields
\begin{equation}
\Gamma_{\rm{UVB}}  = n_{\nu_{\HI}}~c~\bar{\sigma}_{\nu_{\HI}},
\label{eq:Gamma}
\end{equation}
where $n_{\nu_{\HI}}$ is the number density of ionizing photons inside the box. The total number of ionizing photons in the box is therefore given by
\begin{equation}
n_{\nu_{\HI}} L_{\rm{box}}^3 = n_{\gamma}~ 6~N_{\rm{bg}} \frac{L_{\rm{box}}}{c~\Delta t},
\label{eq:tot-phot-box}
\end{equation}
where $n_{\gamma}$ is the number of ionizing photons carried by each
photon packet. Now we can calculate the photon content of each packet
that must be injected into the box during each step in order to achieve the desired $\HI$ photoionization
rate:
\begin{equation}
n_{\gamma} = \frac{\Gamma_{\rm{UVB}}~L_{\rm{box}}^2~\Delta t} {6~\bar{\sigma}_{\nu_{\HI}}~N_{\rm{bg}}},
\label{eq:Gamma}
\end{equation}
\par
We use the redshift-dependent UVB spectrum of \citet{HM01} to calculate $\Gamma_{\rm{UVB}}$ and $\bar{\sigma}_{\nu_{\HI}}$. The \citet{HM01} UVB model successfully reproduces the relative strengths of the observed metal absorption lines in the intergalactic medium \citep{Aguirre08} and has been used to calculate heating/cooling in our cosmological simulations \footnote{Note that during the hydrodynamical simulations, photoheating from the UVB is applied to all gas particles. This ignores the self-shielding of hydrogen atoms against the UVB that occurs at densities $n_{\rm H} \gtrsim 10^{-3}-10^{-2} \cmc$. This inconsistency, which could affect both collisional ionization rates and the small-scale structure of the absorbers, has been found to have no significant impact on the simulated HI CDDF \citep{Pontzen08,McQuinn10,Altay11}. }. 
\par
To reduce the computational cost, we treat the multi-frequency problem in the gray approximation. In other words, we transport the UVB radiation using a single frequency bin, inside which photons are absorbed using the gray cross-section $\bar{\sigma}_{\nu_{\HI}}$ defined in equation \ref{eq:gray-sigma}. Note that the gray approximation ignores the spectral hardening of the radiation field that would occur in multifrequency simulations. In Appendix~\ref{ap:caviets} we show the result of repeating our simulations using multiple frequency bins, and also explicitly accounting for the absorption of photons by helium. These results clearly show the expected spectral hardening. The impact of spectral hardening on the hydrogen neutral fractions and the HI CDDF is small. However, we note that  spectral hardening can change the temperature of the gas in self-shielded regions and that this effect is not captured in our simulations. 
\par
Hydrogen photoionization rates and average absorption cross-sections for UVB radiation at different redshifts are listed in Table \ref{table:UVBz} for our fiducial UVB model based on \citet{HM01} together with \citet{HM12}. The photoionization rate peaks at $z \approx 2-3$ in both models and the equivalent effective photon energy\footnote{We defined the equivalent effective photon energy, $\mathcal{E}_{\rm{eq}}$, which corresponds to the absorption cross section, $\bar{\sigma}_{\nu_{\HI}}$, as: $\mathcal{E}_{\rm{eq}}  \equiv13.6~\rm{eV}\left(\frac{\bar{\sigma}_{\nu_{\HI}}}{\sigma_0}\right)^{-1/3}$ where $\sigma_0 = 6.3 \times 10^{-18} \cms$.} of the background radiation changes only weakly with redshift, compared to the total photoionization rate.

\subsection{Recombination radiation}
\label{sec:recrad}
Photons produced by the recombination of positive ions and electrons can also ionize the gas. If the recombining gas is optically thin, recombination radiation can escape and its ionizing effects can be ignored (i.e., the so-called Case A). However, for regions in which the gas is optically thick, the proper approximation is to assume the ionizing recombination radiation is absorbed on the spot. In this case, the effective recombination rate can be approximated by excluding the transitions that produce ionizing photons \citep[e.g.,][]{Osterbrock06}. This scenario is usually called Case B. A possible way to take into account the effect of recombination radiation is to use Case A recombination at low densities and Case B recombination at high densities \citep[e.g.,][]{Altay11,McQuinn11}, but this will be inaccurate in the transition regime.
\par
In this work we explicitly treat the ionizing photons emitted by recombining hydrogen atoms and follow their propagation through the simulation box. This is facilitated by the fact that the computational cost of RT with TRAPHIC is independent of the number of sources. This is particularly important noting that every SPH particle is potentially a source. The photon production rates of SPH particles depend on their recombination rates and the radiation is emitted isotropically once at the beginning of every RT time step (see Raicevic et al. in prep. for full details).
\par
We do not take into account the redshifting of the recombination photons by peculiar velocities of the emitters, or the Hubble flow. Instead, we assume that all recombination photons are monochromatic with energy $13.6~\rm{eV}$. In reality, recombination photons cannot travel to large cosmological distances without being redshifted to frequencies below the Lyman edge. Therefore, neglecting the cosmological redshifting of RR will result in overestimation of its photoionization rate on large scales. However, because of the small size of our simulation box, the total photoionization rate that is produced by RR on these scales remains negligible compared to the UVB photoionization rate. Consequently, the neglect of RR redshifting is not expected to affect our results.
\begin{figure*}
\centerline{\hbox{\includegraphics[width=0.5\textwidth]
             {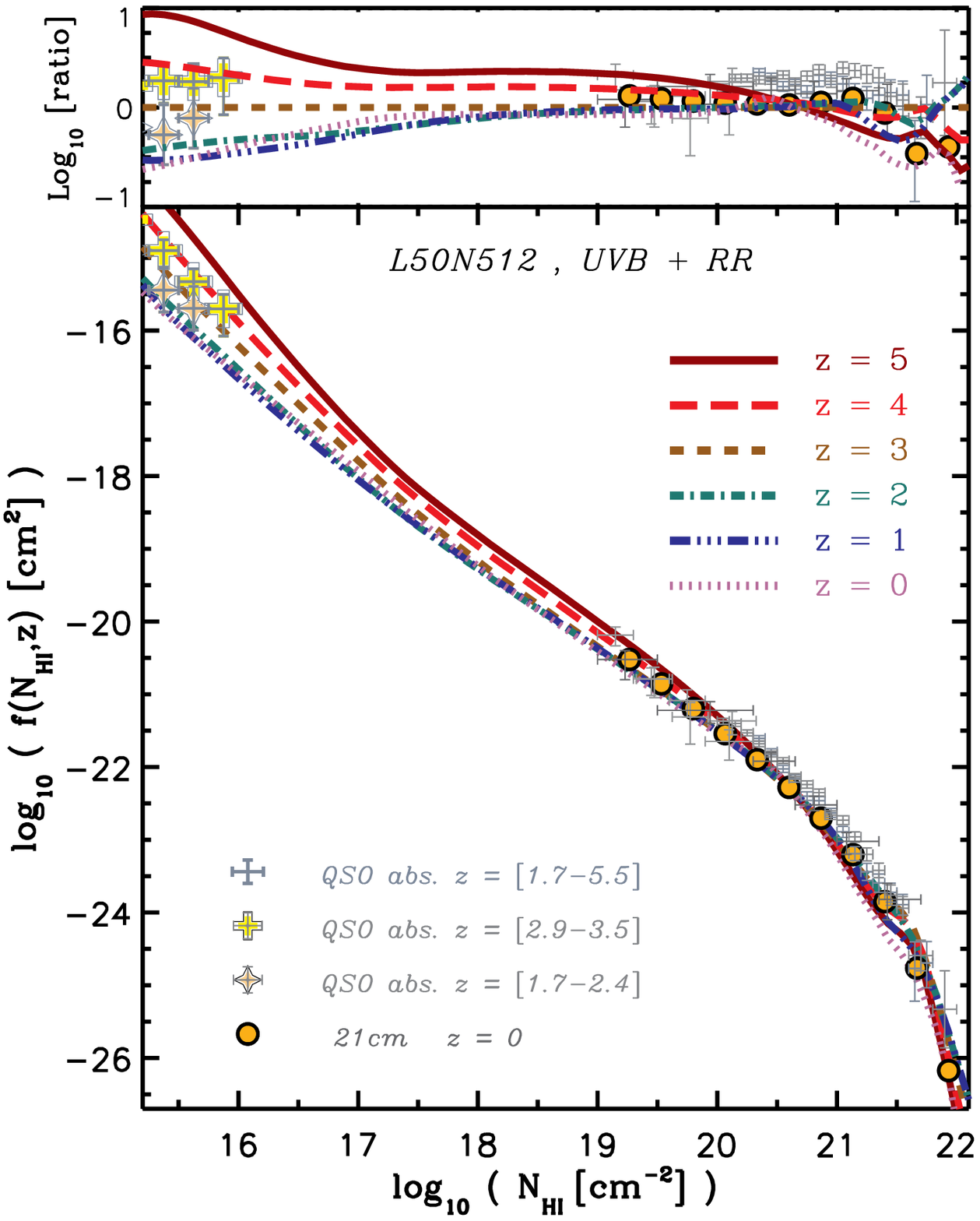}} 
             \hbox{\includegraphics[width=0.5\textwidth]
             {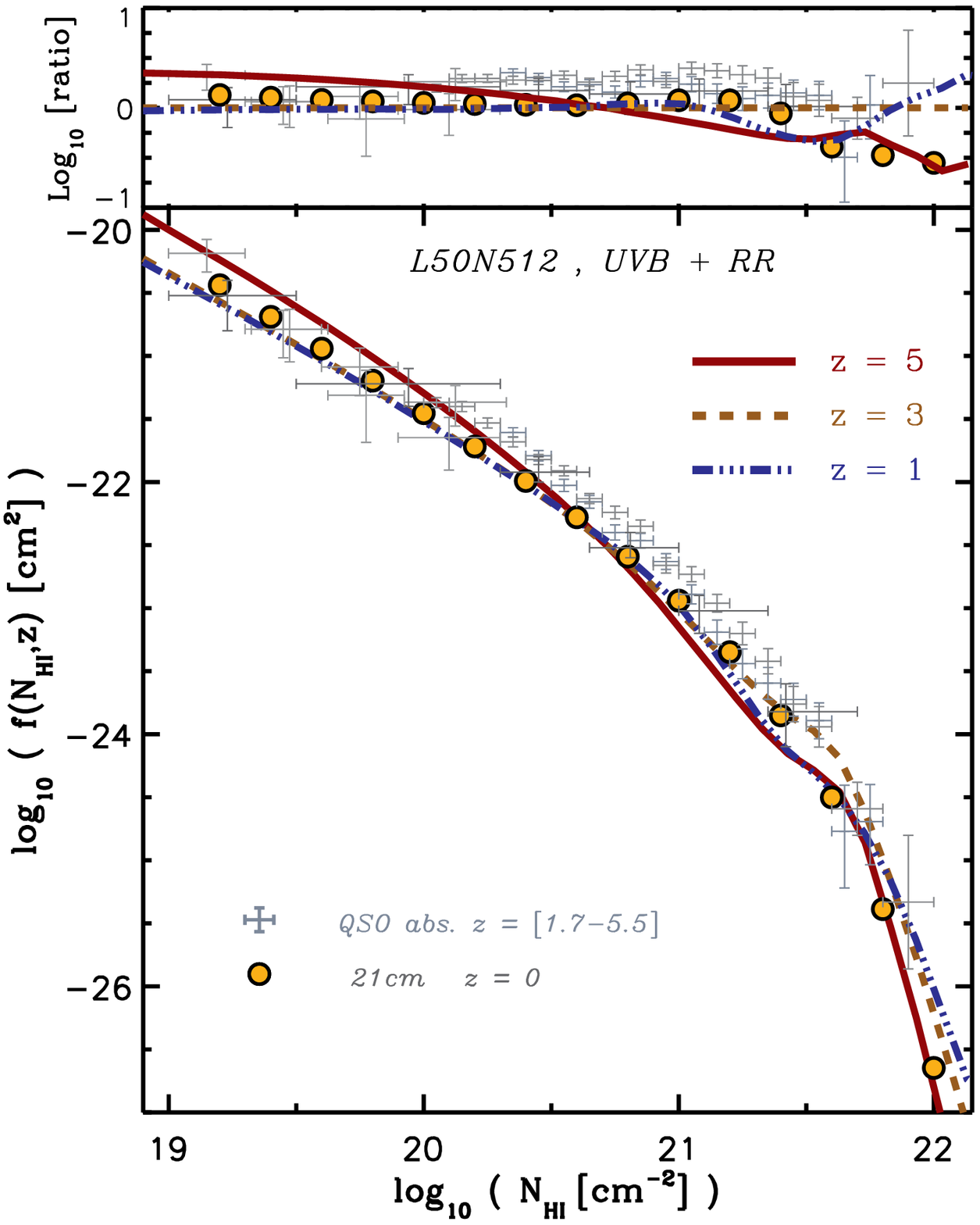}} }
\caption{CDDF of neutral gas at different redshifts in the presence of the UVB and diffuse recombination radiation for \emph{L50N512-W3}. A column density dependent amplitude correction has been applied to make the results consistent with WMAP year 7 cosmological parameters. The observational data points represent a compilation of various quasar absorption line observations at high redshifts (i.e.,  $z = [1.7,5.5]$) taken from \citet{Peroux05} with $z = [1.8,3.5]$, \citet{Omeara07} with $z = [1.7,4.5]$, \citet{Noterdaeme09} with $z = [2.2,5.5]$ and \citet{PW09} with $z = [2.2,5.5]$. The colored data points in the top-left corner of the left panel are taken from \citet{Kim02} with $z = [2.9,3.5]$ and $z = [1.7,2.4]$ for the yellow crosses and orange diamonds, respectively. The orange filled circles show the best-fit based on the low-redshift 21-cm observations of \citet{Zwaan05}. The high column density end of the HI distribution is magnified in the \emph{right panel} and for clarity only the simulated HI CDDF of redshifts $z = 1,~3~\&~5$ are shown. The \emph{top-section} of each panel shows the ratio between the HI CDDFs at different redshifts and the HI CDDF at $z = 3$. The simulation results are in reasonably good agreement with the observations and, like the observations, show only a weak evolution for Lyman Limit and weak damped Ly$\alpha$ systems below $z = 3$.}
\label{fig:column_obs_box}
\end{figure*}

\subsection{The HI column density distribution function}
\label{sec:CDDF}
In order to compare the simulation results with observations, we compute the CDDF of neutral hydrogen, $\fNHI$, which is defined as the number of absorbers per unit column density, per unit absorption length, $d X$:
\begin{eqnarray}
\fNHI \equiv \frac{d^2n}{d \NHI d X}
\equiv \frac{d^2n}{d \NHI d z} \frac{H(z)}{H_0} \frac{1}{(1+z)^2}.
\end{eqnarray}
We project the HI content of the simulation box along each axis onto a grid with $5000^2$ or $10000^2$ pixels (for the $128^3-256^3$ and $512^3$ simulations, respectively)\footnote{Using $5000^2$ cells, the corresponding cell size is similar to the minimum smoothing length, and $\sim 100$ times smaller than the mean smoothing length, of SPH particles at $z = 3$ in the \emph{L06N128} simulation.}. This is done using the actual kernels of SPH particles and for each of the three axes. The projection may merge distinct systems along the line of sight. However, for the small box sizes and high column densities with $\NHI > 10^{17}\cmsq$, which are the focus of this work, the chance of overlap between multiple absorbing systems in projection is negligible. Based on our numerical experiments, we expect that this projection effect starts to appear only at $\NHI < 10^{16}\cmsq$ if one uses a single slice for the projection of the entire \emph{L50N512-W3} simulation box at $z = 3$. To make sure our results are insensitive to this effect, we use, depending on redshift, 25 or 50 slices for projecting the \emph{L50N512-W3} simulation.
\par
To produce a converged $\fNHI$ from simulations, one needs to use cosmological boxes that are large enough to capture the relevant range of over-densities. This is particularly demanding at very high $\HI$ column densities: for instance, \citet{Freeke12} showed that most of the gas with $\NHI > 10^{21} \cmsq$ resides in galaxies with halo masses $\gtrsim 10^{11}~\Msun$ which are relatively rare. As we show in Appendix \ref{ap:res-box-cos-tests}, the box size required to produce a converged $\HI$ CDDF up to $\NHI \sim 10^{22} \cmsq$ is $L \gtrsim 50$ comoving $\Mpch$. Simulating RT in such a large volume is expensive. However, as we show in \S\ref{sec:result-z3}, at a given redshift the photoionization rates are fit very well by a function of the hydrogen number density. This relation is conserved with respect to both box size\footnote{One should note that the box size can indirectly change the resulting photoionization rate profile. For instance, self-shielding can be affected by collisional ionization, which become stronger at lower redshifts and whose importance depends on the abundance of massive objects, which is more sensitive to the box size.} and resolution and can therefore be applied to our highest resolution simulation (i.e., \emph{L50N512-W3}), allowing us to keep the numerical cost tractable. Finally, since repeating the high-resolution simulations is expensive, we apply a redshift-independent correction which accounts for the difference between the WMAP year 3 parameters used for \emph{L50N512-W3} simulation and the WMAP year 7 values. This is done by multiplying all the $\HI$ CDDFs produced based on the WMAP year 3 cosmology by the ratio between the $\HI$ CDDFs for \emph{L25N512} and \emph{L25N512-W3} at $z =3$.

\subsection{Dust and molecular hydrogen}
\label{sec:dust-H2}
Dust and star formation are highly correlated and infrared observations indicate that the prevalence of dusty galaxies follows the average star formation history of the Universe \citep[e.g.,][]{Rahmati11}. Nevertheless, dust extinction is a physical processes that is not treated in our simulations. Assuming a constant dust-to-gas ratio, the typical dust absorption cross-section per atom is orders of magnitudes lower than the typical hydrogen absorption cross-section for ionizing photons \citep{Weingartner01}. In other words, the absorption of ionizing photons by dust particles is not significant compared to the absorption by the neutral hydrogen. Consequently, as also found in cosmological simulations with ionizing radiation \citep{Gnedin08}, dust absorption does not noticeably alter the overall distribution of ionizing photons and hydrogen neutral fractions. 
\par
The observed cut-off in the abundance of very high $\NHI$ systems may be related to the conversion of atomic hydrogen into $\rm{H}_2$ \citep[e.g.,][]{Schaye01b, Krumholz09,PW09,Altay11}. Following \citet{Altay11} and \citet{Duffy12}, we adopt an observationally driven scaling relation between gas pressure and hydrogen molecular fraction \citep{Blitz06} in post-processing, which reduces the amount of observable $\HI$ at high densities. This scaling relation is based on observations of low-redshift galaxies and may not cover the low metallicities relevant for higher redshifts. This could be an issue, since the $\HI$-$\Hm$ relation is known to be sensitive to the dust content and hence to the metallicity \citep[e.g.,][]{Schaye01b,Schaye04,Krumholz09a}. 
\section{Results}
\label{sec:results}
In this section we report our findings based on various RT simulations which include UVB ionizing radiation and diffuse recombination radiation from ionized gas. As we demonstrate in \S\ref{sec:Photoionization-density-fit}, the dependence of the photoionization rate on density obtained from our RT simulations shows a generic trend for different resolutions and box sizes. Therefore, we can use the results of RT calculations obtained from smaller boxes (e.g., \emph{L06N128} or  \emph{L06N256}) which are computationally cheaper, to calculate the neutral hydrogen distribution in larger boxes. The last column of Table \ref{table:hydro_sims} indicates for which simulations this was done.
\par
In the following, we will first present the predicted HI CDDF and compare it with observations. Next we discuss other aspects of our RT results and the effects of ionization by the UVB, recombination radiation and collisional ionization  on the resulting HI distributions at different redshifts.

\subsection{Comparison with observations}

In Figure \ref{fig:column_obs_box} we compare the simulation results with a compilation of observed $\HI$ CDDFs, after converting both to the WMAP year 7 cosmology. The data points with error bars show results from high-redshift ($z = 1.7 - 5.5$) QSO absorption line studies and the orange filled circles show the fitting function reported by \citet{Zwaan05} based on 21-cm observations of nearby galaxies. The latter observations only probe column densities $\NHI \gtrsim 10^{19}~\cmsq$.
\par
We note that the OWLS simulations have already been shown to agree with observations by \citet{Altay11}, but only for $z = 3$ and based on a different RT method (see Appendix \ref{ap:Altay-comp} for a comparison). Overall, our RT results are also in good agreement with the observations. At high column densities (i.e., $\NHI > 10^{17}~\cmsq$) the observations probing $0 <z <5.5$ are consistent with each other. This implies weak or no evolution with redshift. The simulation is consistent with this remarkable observational result, predicting only weak evolution for $ 10^{17} \cmsq < \NHI <10^{21} \cmsq $ (i.e., Lyman limit systems, LLSÕs, and weak Damped Ly-$\alpha$ systems, DLAÕs) especially at $z \lesssim 3$. 
\par
The simulation predicts some variation with redshift for strong DLAÕs ($\NHI \gtrsim 10^{21} \cmsq $). The abundance of strong DLAÕs in the simulations follows a similar redshift-dependent trend as the average star formation density in our simulations which peaks at $z \approx 2-3$ \citep{Schaye10}. This result is consistent with the DLA evolution found by \citet{Cen12} in two zoomed simulations of a cluster and a void. One should, however, note that at very high column densities (e.g., $\NHI \gtrsim 10^{21.5} \cmsq $) both observations and simulations are limited by small number statistics and the simulation results are more sensitive to the adopted feedback scheme (Altay et al. in prep.). Moreover, as we will show in Rahmati et al. (in prep.), including local stellar ionizing radiation can decrease the $\HI$ CDDF by up to $\approx 1$ dex for $\NHI \gtrsim 10^{21}~\cmsq$, especially at redshifts $z \approx 2-3$ for which the average star formation activity of the Universe is near its peak. 
 \par
At low column densities (i.e., $\NHI \lesssim 10^{17}~\cmsq$) the simulation results agree very well with the observations. This is apparent from the agreement between the simulated $\fNHI$ at $z = 3$  and $z = 4$, and the observed values for redshifts $2.9 < z < 3.5$ \citep{Kim02} which are shown by the yellow crosses in the left panel of Figure \ref{fig:column_obs_box}. The simulated $\fNHI$ at lower and higher redshifts deviate from those at $z \approx 3$ showing the abundance of those systems decreases with decreasing redshift and remains nearly constant at $z \lesssim 2$. This is consistent with the Ly-$\alpha$ forest observations at lower redshifts \citep{Kim02,Janknecht06,Lehner07,PW09,Ribaudo11}, as illustrated with the orange diamonds which correspond to $z \approx 2$ observations, in the top-left corner of the left panel in Figure \ref{fig:column_obs_box}. 
\par
The evolution of the $\HI$ CDDF with redshift results from a combination of the expanding Universe and the growing intensity of the UVB radiation down to redshifts $z \approx 2-3$. At low redshifts (i.e., $z \approx 0$) the intensity of the UVB radiation has dropped by more than one order of magnitude leading to higher hydrogen neutral fractions and higher $\HI$ column densities. However, as we show in \S\ref{sec:result-zs}, at lower redshifts an increasing fraction of low-density gas is shock-heated to temperatures sufficiently high to become collisionally ionized and this compensates for the weaker UVB radiation at low redshifts.
\par
The simulated $\HI$ CDDFs at all redshifts are consistent with each other and the observations. However, as illustrated in the right panel of Figure \ref{fig:column_obs_box}, there is a $\approx 0.2$ dex difference between the simulation results and the observations of LLS and DLAs at all redshifts. We found that the normalization of the $\HI$ CDDF in those regimes is sensitive to the adopted cosmological parameters (see Appendix \ref{ap:res-box-cos-tests}). Notably, the cosmology consistent with the WMAP 7 year results that is shown here, produces a better match to the observations than a cosmology based on the WMAP 3 year results with smaller values for $\Ob$ and $\sigeight$. This suggests that a higher value of $\sigeight$ may explain the small discrepancy between the simulation results and the observations.
\begin{figure*}
\centerline{\hbox{\includegraphics[width=0.45\textwidth]
             {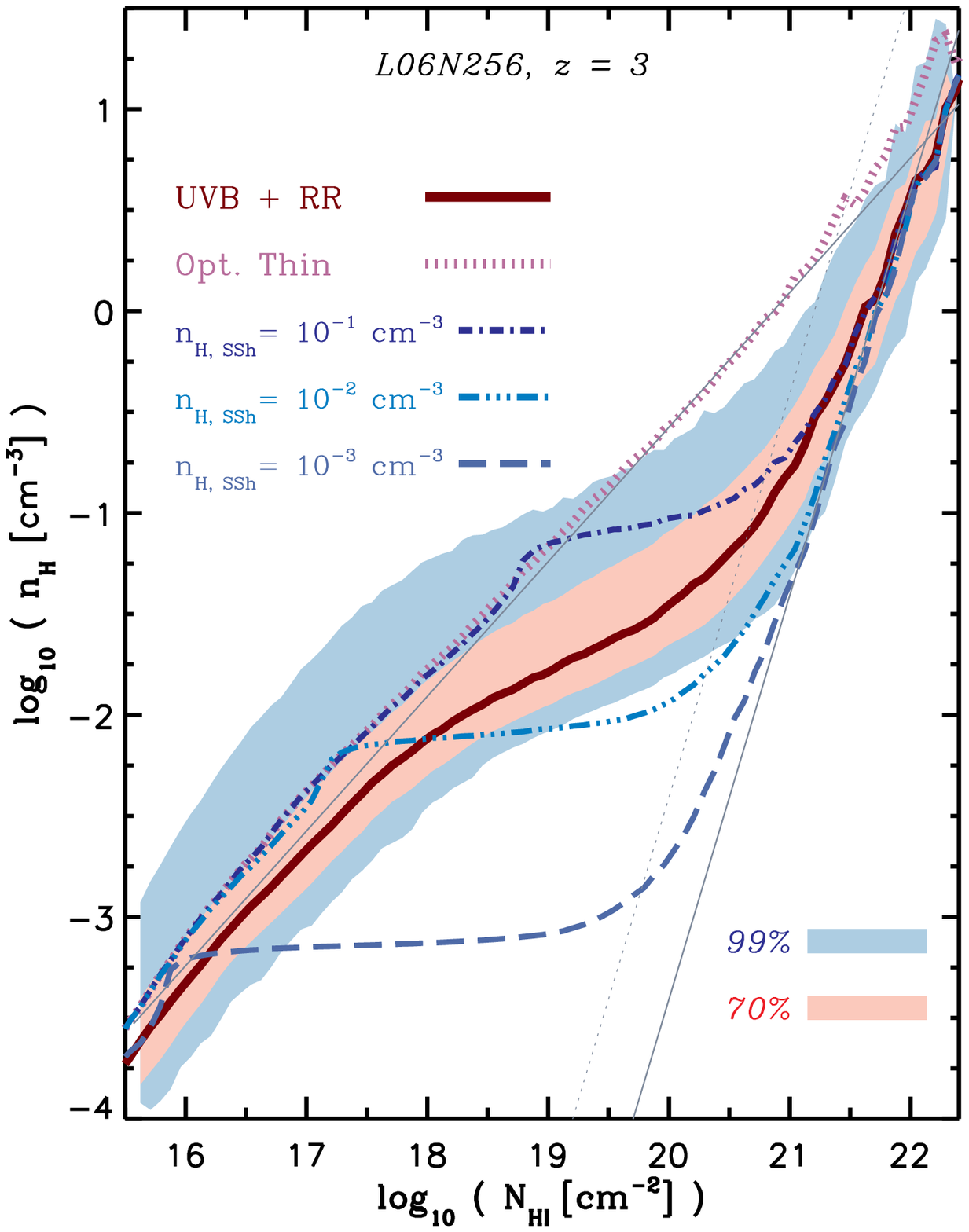}} 
             \hbox{\includegraphics[width=0.45\textwidth]
             {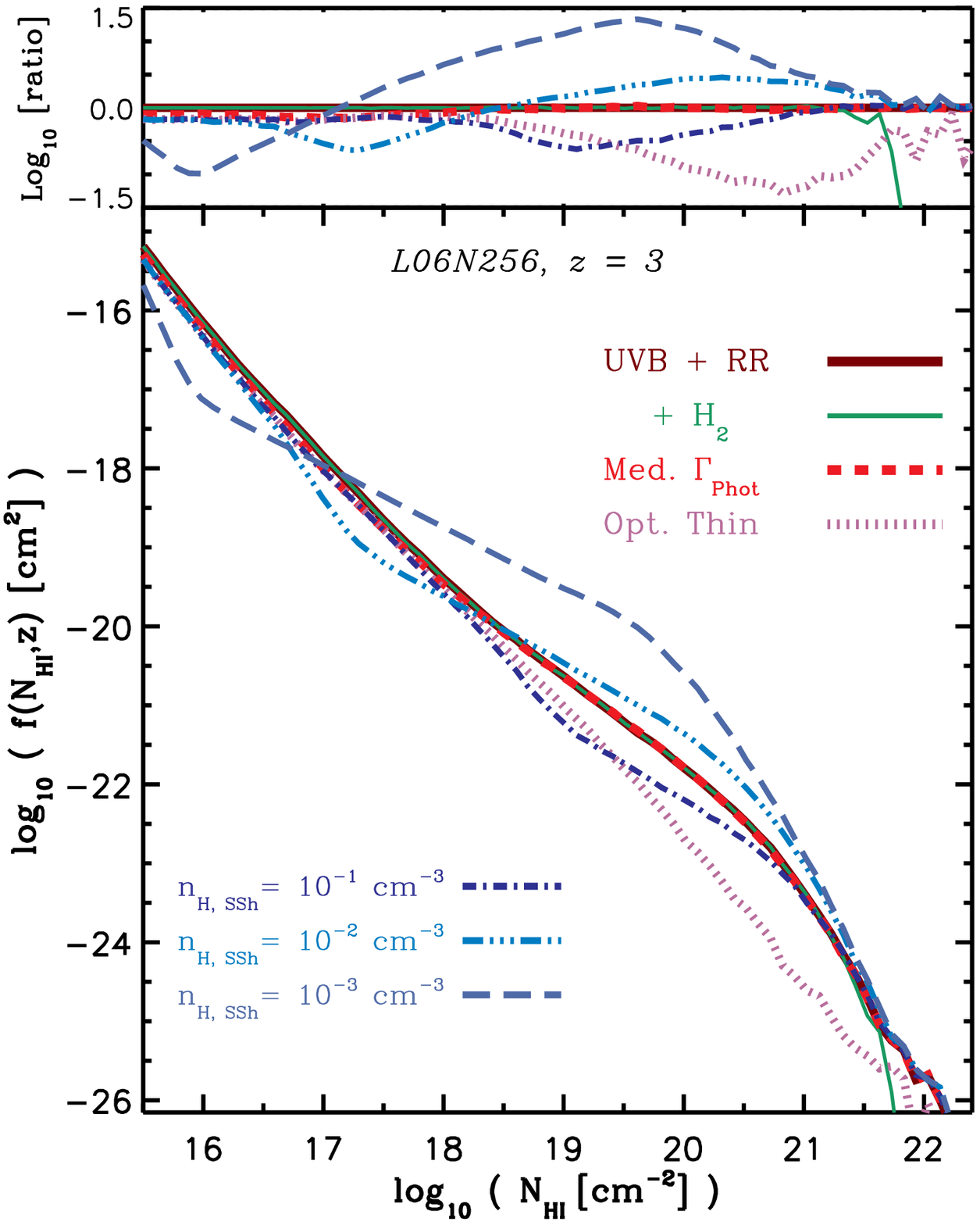}} }
\caption{ \emph{Left}: $\nHI$-weighted total hydrogen number density as a function of $\NHI$. The brown solid curve shows the RT results and the purple dotted curve shows the optically thin limit. Blue dot-dashed, dot-dot-dot-dashed and long dashed curves assume models with self-shielding density thresholds of $n_{\rm{H, SSh}}= 10^{-1},~ 10^{-2}~$ and $10^{-3} \cmcb$, respectively. All of the above mentioned curves show the median of  the $\nHI$-weighted total hydrogen number density at a given $\NHI$. The gray thin lines show the expected Jeans scaling relations for optically thin gas (equation \ref{eq:ToTden-col-density}; diagonal solid line) and for neutral gas (equation \ref{eq:den-col-density}; steeper dotted line). A second solid line with the same slope expected from equation \eqref{eq:den-col-density} but a different normalization is illustrated by the second solid line which is identical to the dotted line but shifted by 0.5 dex to higher $\NHI$. The pink and blue shaded areas in the right panel indicate the $70\%$ and $99\%$ scatter respectively, while the solid curves shows the median for the RT result. All the other curves are also medians. This shows that the $\nH - \NHI$ relationship can be explained by the Jeans scaling and that the flattening in the CDDF is due to self-shielding.
\emph{Right}: HI CDDF in the presence of the UVB and diffuse recombination radiation for simulation \emph{L06N256}. Simulations shown with different curves are identical to the left panel. In addition, the effect of $\Hm$ formation is shown by the green solid curve which deviates from the brown solid curve at $\NHI \gtrsim 3 \times 10^{21}\cmsq$. Finally the red dashed curve, which is indistinguishable from the brown solid curve, shows the result of assuming the median of the photoionization rate profile of the RT results to calculate the neutral fractions (see $\S$\ref{sec:Photoionization-density-fit} and Appendix \ref{ap:RT-fit-tests}). The \emph{top-section} in the right panel shows the ratio between different HI CDDFs and the one resulting from the RT simulations. }
\label{fig:column_nHI_flat}
\end{figure*}

\subsection{The shape of the HI CDDF}
\label{sec:fNHI-shape}
The shape of the HI CDDF is determined by the distribution of hydrogen and by the different ionizing processes that set the hydrogen neutral fractions of the absorbers. One can assume that over-dense hydrogen resides in self-gravitating systems that are in local hydrostatic equilibrium. Then, the typical scales of the systems can be calculated as a function of the gas density based on a Jeans scaling argument \citep{Schaye01}. Assuming that absorbers have universal baryon fractions (i.e., $f_{\rm{g}} = \frac{\Ob}{\Om}$) and typical temperatures of $T_4 \equiv (T/ 10^4~K)\sim1$ (i.e., collisional ionization is unimportant), one can calculate the total hydrogen column density \citep{Schaye01}:
\begin{equation}
 N_{\rm H} \sim~  1.6~\times~10^{21}~ \cmsq~ \nH^{1/2}  ~T^{1/2}_{4}~ \left(\frac{f_{\rm{g}}}{0.17}\right)^{1/2}.
\label{eq:ToTden-col-density}
\end{equation}
Assuming that the gas is highly ionized and in ionization equilibrium with the ambient ionizing radiation field with the photoionization rate, $\Gamma_{-12} = \Gamma / 10^{-12}~\rm{s^{-1}}$, one gets \citep{Schaye01}:
\begin{eqnarray}
&{}& \NHI \sim~  2.3~\times~10^{13}~ \cmsq \left(\frac{\nH}{10^{-5}~\cmcb}\right)^{3/2} \nonumber \\
&{}& \qquad  \qquad  \times ~ T^{-0.26}_{4}~ \Gamma_{-12}^{-1}~\left(\frac{f_{\rm{g}}}{0.17}\right)^{1/2}.
\label{eq:den-col-density}
\end{eqnarray}
\par
At high densities where the gas is nearly neutral, equation \eqref{eq:ToTden-col-density} provides a relation between $\NHI$ and $\nH$. Equation \eqref{eq:den-col-density} on the other hand, gives the relation for optically thin, highly ionized gas. The latter is derived assuming that the UVB photoionization is the dominant source of ionization, which is a good assumption at high redshifts and explains the relation between density and column density in Ly$\alpha$ forest simulations \citep[e.g.,][]{Dove10, Altay11, McQuinn11,Tepper12}. However, as we will show in the following sections, photoionization domination breaks down at lower redshifts where collisional ionization plays a significant role. 
\par
The column density at which hydrogen starts to be self-shielded against the UVB radiation follows from setting $\tau_{\HI} = 1$:
\begin{equation}
 \NHI_{\rm{,SSh}} \sim 4~\times~10^{17}~ \cmsq \left(\frac{\bar{\sigma}_{\nu_{\HI}}}{2.49 \times 10^{-18}~\cms}\right)^{-1}
\label{eq:Opt-depth1}
\end{equation}
which can be used together with equation \eqref{eq:den-col-density} to find the typical densities at which the self-shielding begins \citep[e.g.,][]{Furlanetto05}:
\begin{eqnarray}
&{}& {n_{\rm{H,SSh}}} \sim~  6.73\times10^{-3} \cmcb \left(\frac{\bar{\sigma}_{\nu_{\HI}} }{2.49\times10^{-18}\cms}\right)^{-2/3} \nonumber \\
&{}& \qquad  \qquad  \times ~ T^{0.17}_{4}~ \Gamma_{-12}^{2/3}~\left(\frac{f_{\rm{g}}}{0.17}\right)^{-1/3}.
\label{eq:densitySSH}
\end{eqnarray}
\par
These relations are compared with the $\nHI$-weighted total hydrogen number density as a function of $\NHI$ in the \emph{L06N256} simulation at $z = 3$ in the left panel of Figure \ref{fig:column_nHI_flat}. The solid curve shows the median and the red (blue) shaded area represents the central $70\%$ ($90\%$) percentile. The diagonal gray solid line which converges with the simulation results at low column densities, shows equation \eqref{eq:den-col-density} and the steeper gray dotted line which converges with the simulation results at high column densities is based on equation \eqref{eq:ToTden-col-density}. The agreement between the expected slopes of the $\nH - \NHI$ relation and the simulations at low and high column densities confirms our initial assumption that hydrogen resides in self-gravitating systems which are close to local hydrostatic equilibrium\footnote{One should note that the above mentioned Jeans argument provides an order of magnitude calculation due to its simplifying assumptions (e.g., uniform density, universal baryon fraction, etc.). Although we may expect the predicted scaling relations to be correct, the very close agreement of the normalization with the simulations at low densities is coincidental. As the steeper gray dotted line which is based on equation \eqref{eq:ToTden-col-density} shows, the simulated $\NHI$ for a given $\nH$ is $\approx 0.5$ dex higher than implied by the Jeans scaling for the nearly neutral case (i.e., steep, gray solid line).}. 
\par
As expected from equation \eqref{eq:densitySSH}, at low densities the gas is optically thin and follows the Jeans scaling relation of the highly ionized gas. At $\nH~\gtrsim~ 0.01\cmcb$ however, the relation between density and column density starts to deviate from equation \eqref{eq:den-col-density} and approaches that of a nearly neutral gas. Consequently, for densities above the self-shielding threshold the $\HI$ column density increases rapidly over a narrow range of densities, leading to a flattening in the $\nH - \NHI$ relation and in the resulting $\HI$ CDDF at $\NHI \gtrsim 10^{18}\cmsq$ (see Figure \ref{fig:column_nHI_flat}). The results from the RT simulation deviate from the magenta dotted lines, which are obtained assuming optically thin gas, at $\NHI \gtrsim 4\times10^{17}\cmsq$. As the dotted line in the right panel of Figure \ref{fig:column_nHI_flat} shows, in the absence of self-shielding, the slope of $\fNHI \propto \NHI^{\beta}$ is constant all the way up to DLAs at $\beta_{\rm{Ly}\alpha} \approx -1.6$. However, because of self-shielding, the $\HI$ CDDF flattens to $\beta_{\rm{LLS}} \approx -1.1$ at $10^{18}\cmsq \lesssim \NHI \lesssim 10^{20}\cmsq$ in the RT simulation (solid curve). These predicted slopes are in excellent agreement with the latest observational constraints of $\beta_{\rm{Ly}\alpha} \gtrsim -1.6$ for $10^{15}\cmsq < \NHI <10^{17}\cmsq$ to $\beta_{\rm{LLS}} \approx -1$ in the LLS regime \citep{Omeara12}. We also note that $\beta_{\rm{Ly}\alpha} \gtrsim -1.6$ is predicted to be almost the same for all redshifts, which agrees well with observations \citep{Janknecht06,Lehner07,Ribaudo11}.
\par
At densities $\nH \gtrsim 0.1\cmcb$ the gas is nearly neutral and the Jeans scaling in equation \eqref{eq:ToTden-col-density} controls the $\nH - \NHI$ relation. Consequently, the rate at which $\NHI$ responds to changes in $\nH$ slows down, causing a steepening in the resulting $\fNHI$ in the DLA range (i.e., $\NHI \gtrsim 10^{21} \cmsq$). However, as the thick solid curve in the right panel of Figure \ref{fig:column_nHI_flat} illustrates, the slope of $\fNHI$ remains constant for $\NHI = 10^{21} - 10^{22}~ \cmsq$. This is in contrast with observed trends indicating a sharp cut-off at $\NHI \gtrsim 3 \times 10^{21} \cmsq$ (\citealp{Prochaska10, Omeara12}; but see \citealp{Noterdaeme12}). At those column densities a large fraction of hydrogen is expected to form $\rm{H_2}$ molecules and be absent from $\HI$ observations \citep{Schaye01b, Krumholz09,Altay11}. As the thin solid line in the right panel of Figure \ref{fig:column_nHI_flat} shows, accounting for $\rm{H_2}$ using the empirical relation between $\rm{H_2}$ fraction and pressure, based on $z = 0$ observations \citep{Blitz06}, does reproduce a sharp cut-off. If the observed relation does not cut off \citep{Noterdaeme12}, then this may imply that $\rm{H_2}$ fractions are lower at $z = 3$ than at $z = 0$. We also note that the ionizing effect of local sources (Rahmati et al. in prep.), increasing the efficiency of stellar feedback, e.g.,  by using a top-heavy IMF, and AGN feedback can also affect these high $\HI$ column densities (Altay et al. in prep.).
\begin{figure*}
\centerline{\hbox{\includegraphics[width=0.45\textwidth]
             {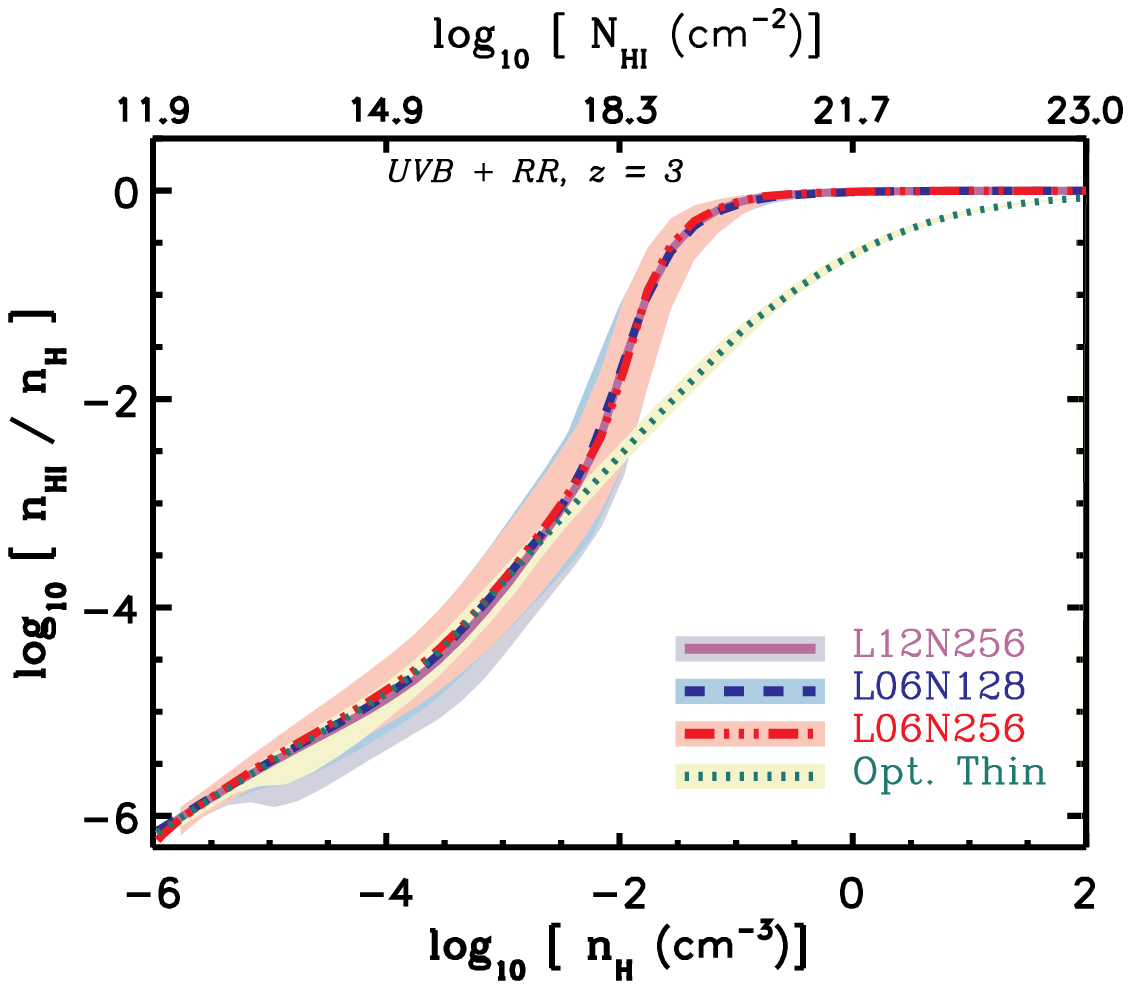}} 
             \hbox{\includegraphics[width=0.45\textwidth]
             {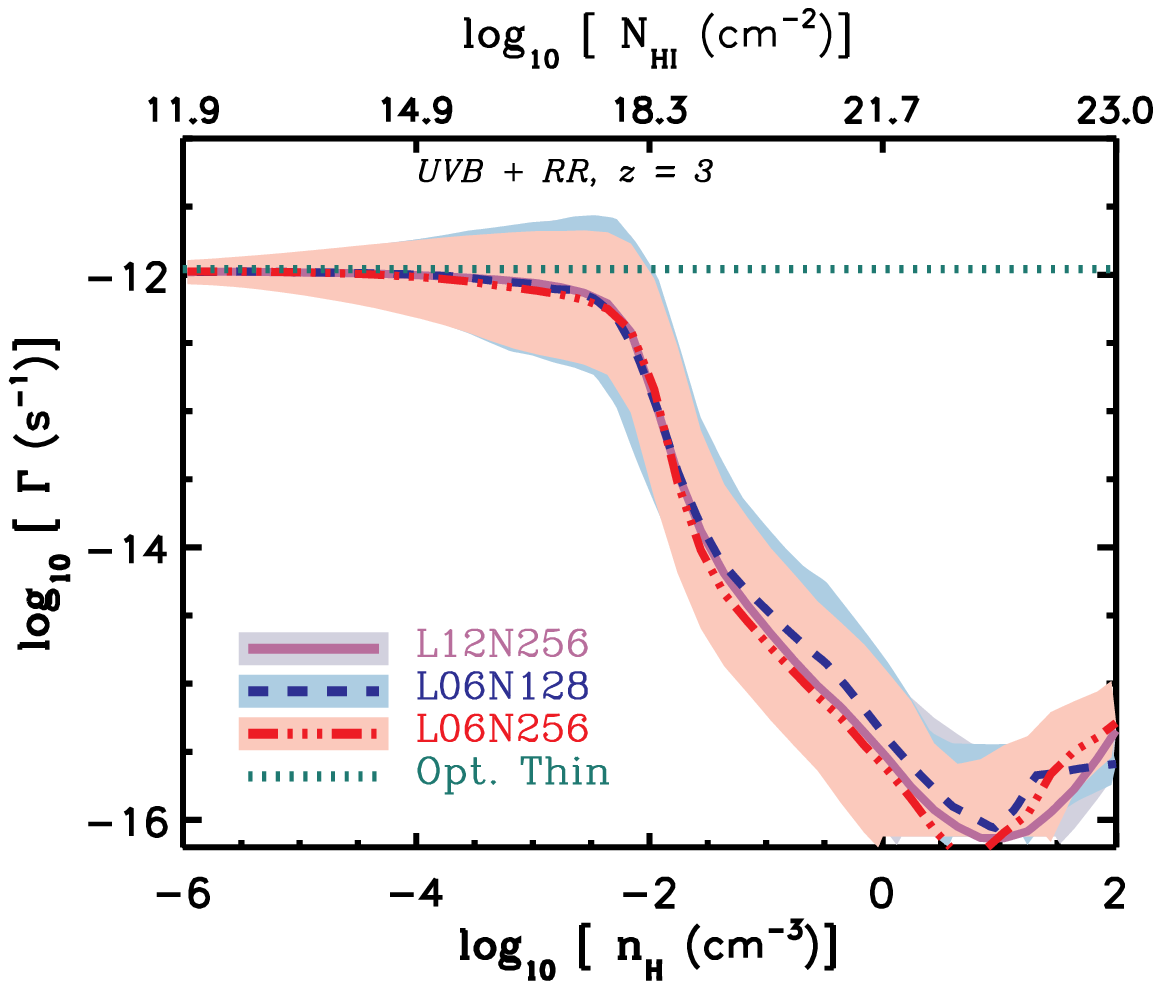}} }
\caption{The hydrogen neutral fraction (\emph{left}) and the photoionization rate (\emph{right}) as a function of hydrogen number density do not change by varying the simulation box size or mass resolution. This is shown for different simulations at z = 3 in the presence of the UVB and recombination radiation. Purple solid, blue dashed and red dot-dashed lines show, respectively, the results for \emph{L12N256}, \emph{L06N128} and \emph{L06N256}. The green dotted line indicates the results for the \emph{L06N128} simulation if the gas is assumed to be optically thin to the UVB radiation (i.e., no RT calculation is performed). The deviation between the optically thin hydrogen neutral fractions and RT results at $\nH \gtrsim10^{-2}\cmcb$ shows the impact of self-shielding. The lines show the medians and the shaded areas indicate the $15\% - 85\%$ percentiles. At the top of each panel we show HI column densities corresponding to each density. }
\label{fig:Gamma_eta_Res}
\end{figure*}
\par
To first order, one can mimic the effect of RT by assuming gas with $\nH < n_{\rm{H,SSh}}$ to be optically thin (i.e., Case A recombination) and gas with $\nH > n_{\rm{H,SSh}}$ to be fully neutral. Simulations with three different self-shielding density thresholds are shown in Figure \ref{fig:column_nHI_flat}. The dot-dashed, dot-dot-dot-dashed and long dashed curves correspond to $n_{\rm{H, SSh}}= 10^{-1},~ 10^{-2}$ and $10^{-3} \cmcb$, respectively. Although all of these simulations predict the flattening of $\fNHI$, they produce a transition between optically thin and neutral gas that is too steep. In contrast, the RT results show a transition between highly ionized and highly neutral gas that is more gradual, as observed.
\subsection{Photoionization rate as a function of density}
\label{sec:Photoionization-density-fit}
\begin{figure}
\centerline{\hbox{\includegraphics[width=0.45\textwidth]
             {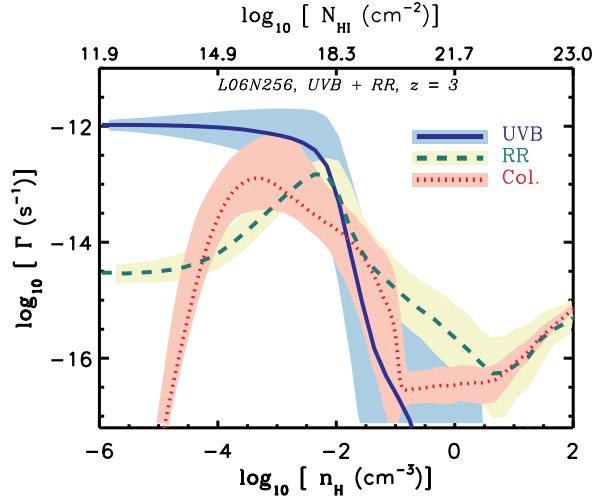}} }
\caption{Ionization rates due to different sources of ionization as a function of hydrogen number density. Blue solid, green dashed and red dotted curves show, respectively, the UVB photoionization rate, the recombination radiation photoionization rate and the collisional ionization rate. The curves show the medians and the shaded areas around the medians indicate the $15\% - 85\%$ percentiles. HI column densities corresponding to each density are shown along the top x-axis. While the UVB is the dominant source of ionization below the self-shielding (i.e., $\nH \lesssim 10^{-2}\cmcb$), recombination radiation dominates the ionization at higher densities.}
\label{fig:Gamma_z3_comp}
\end{figure}

Figure \ref{fig:Gamma_eta_Res} illustrates the RT results for neutral fractions and photoionization rates as a function of density in the presence of UVB radiation and diffuse recombination radiation for the \emph{L06N128}, \emph{L06N256} and \emph{L12N256} simulations at z = 3. For comparison, the results for the optically thin limit are shown by the green dotted curves. The sharp transition between highly ionized and neutral gas and its deviation from the optically thin case are evident in the left panel. This transition can also be seen in the photoionization rate (right panel) which drops at $\nH \gtrsim 0.01~\cmcb$, consistent with equation \eqref{eq:densitySSH} and previous studies \citep{Tajiri98, Razoumov06, Faucher10, Nagamine10, Fumangalli11, Altay11}. 
\par
The medians and the scatter around them are insensitive to the resolution of the underlying simulation and to the box size. This suggests that one can use the photoionization rate profile obtained from the RT simulations for calculating the hydrogen neutral fractions in other simulations for which no RT has been performed.
\par
Moreover, as we show in $\S$\ref{sec:result-zs}, the total photoionization rate as a function of the hydrogen number density has the same shape at different redshifts. This shape can be characterized by three features: i) a knee at densities around the self-shielding density threshold, ii) a relatively steep fall-off at densities higher than the self-shielding threshold and iii) a flattening in the fall-off after the photoionization rate has dropped by $\sim 2$ dex from its maximum value which is caused by the RR photoionization. These features are captured by the following fitting formula:
\begin{eqnarray}
&{}&\frac{\Gamma_{\rm{Phot}}} {\Gamma_{\rm{UVB}}} = 0.98~\left[1 + \left(\frac{\nH}{n_{\rm{H, SSh}}}\right)^{1.64}  \right]^{-2.28} \nonumber \\
&{}& \qquad  \qquad \qquad  \qquad \qquad +0.02~\left[1 + \frac{\nH}{n_{\rm{H, SSh}}} \right]^{-0.84},
\label{eq:Gamma-fit}
\end{eqnarray}
where $\Gamma_{\rm{UVB}}$ is the background photoionization rate and $\Gamma_{\rm{Phot}}$ is the total photoionization rate. Moreover, the self-shielding density threshold, $n_{\rm{H, SSh}}$, is given by equation \eqref{eq:densitySSH} and is thus a function of $\Gamma_{\rm{UVB}}$ and $\bar{\sigma}_{\nu_{\HI}}$ which vary with redshift. As explained in more detail in Appendix \ref{ap:RT-fit-tests}, the numerical parameters representing the shape of the profile are chosen to provide a redshift independent best fit to our RT results. In addition, the parametrization is based on the main RT related quantities, namely the intensity of UVB radiation and its spectral shape. It can therefore be used for UVB models similar to the \citet{HM01} model we used in this work \citep[e.g.,][]{Faucher09,HM12}. For a given UVB model, one only needs to know $\Gamma_{\rm{UVB}}$ and $\bar{\sigma}_{\nu_{\HI}}$ in order to determine the corresponding $n_{\rm{H, SSh}}$ from \eqref{eq:densitySSH} (see also Table \ref{table:UVBz}). Then, after using equation \eqref{eq:Gamma-fit} to calculate the photoionization rate as a function of density, the equilibrium hydrogen neutral fraction for different densities, temperatures and redshifts can be readily calculated as explained in Appendix \ref{ap:RT-fit-tests}.
\par
We note that the parameters used in equation \eqref{eq:Gamma-fit} are only accurate for photoionization dominated cases. As we show in \S\ref{sec:result-zs}, at $z \sim 0$ the collisional ionization rate is greater than the total photoionization rate around the self-shielding density threshold. Consequently, equation \eqref{eq:densitySSH} does not provide an accurate estimate of the self-shielding density threshold at low redshifts. In Appendix \ref{ap:RT-fit-tests} we therefore report the parameters that best reproduce our RT results at $z = 0$. Our tests show that simulations that use equation \eqref{eq:Gamma-fit} reproduce the $\fNHI$ accurately to within $10\%$ for $z \gtrsim 1$ where photoionization is dominant (see Appendix \ref{ap:RT-fit-tests}).
\par
Although using the relation between the median photoionization rate and the gas density is a computationally efficient way of calculating equilibrium neutral fractions in big simulations, it comes at the expense of the information encoded in the scatter around the median photoionization rate at a given density. However, our experiments show that the error in $\fNHI$ that results from neglecting the scatter in the photoionization rate profile is negligible for $\NHI \gtrsim 10^{18}\cmcb$ and less than $\lesssim 0.1$ dex at lower column densities (see Appendix \ref{ap:RT-fit-tests}). 
\par
\subsection{The roles of diffuse recombination radiation and collisional ionization at $z = 3$}
\label{sec:result-z3}
\begin{figure*}
\centerline{\hbox{\includegraphics[width=0.45\textwidth]
             {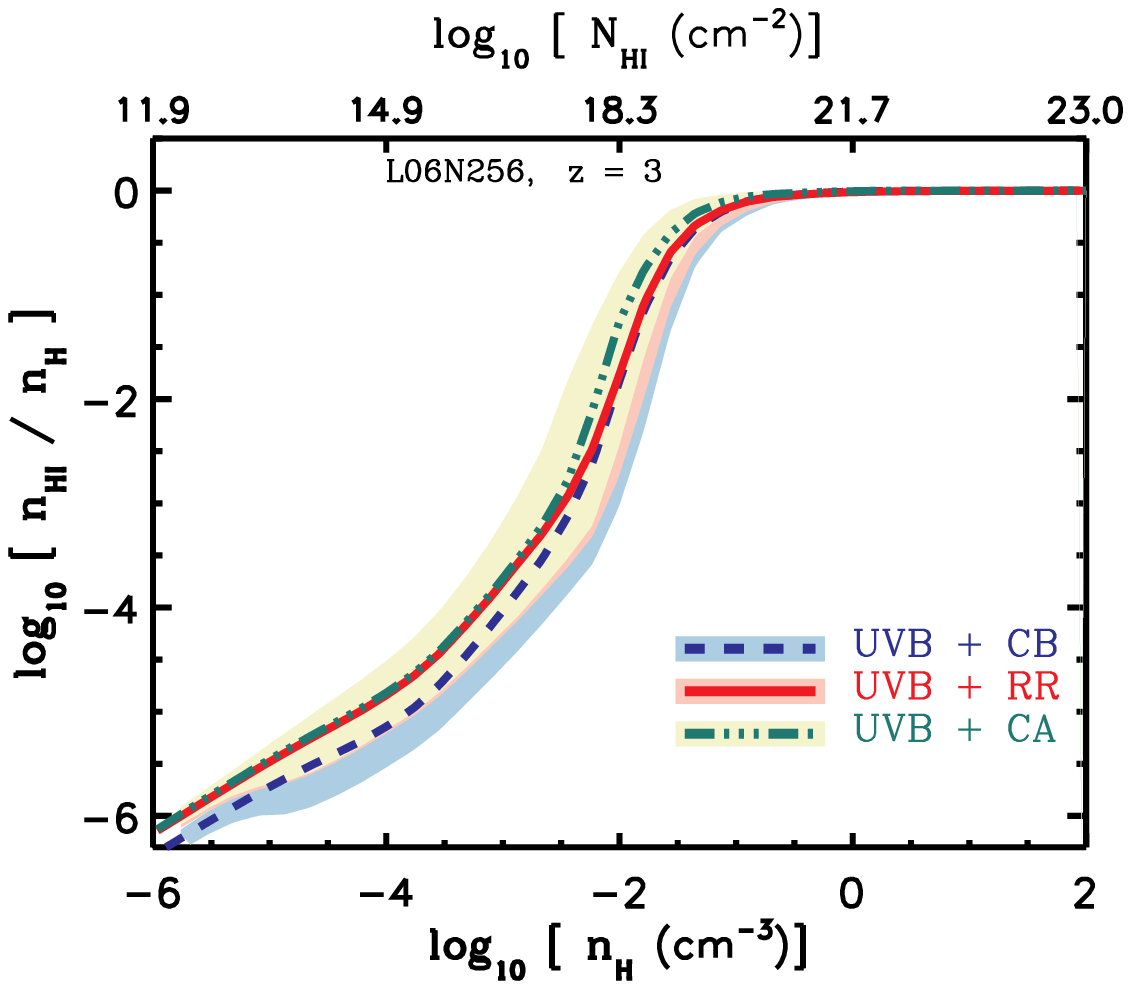}} 
             \hbox{\includegraphics[width=0.45\textwidth]
             {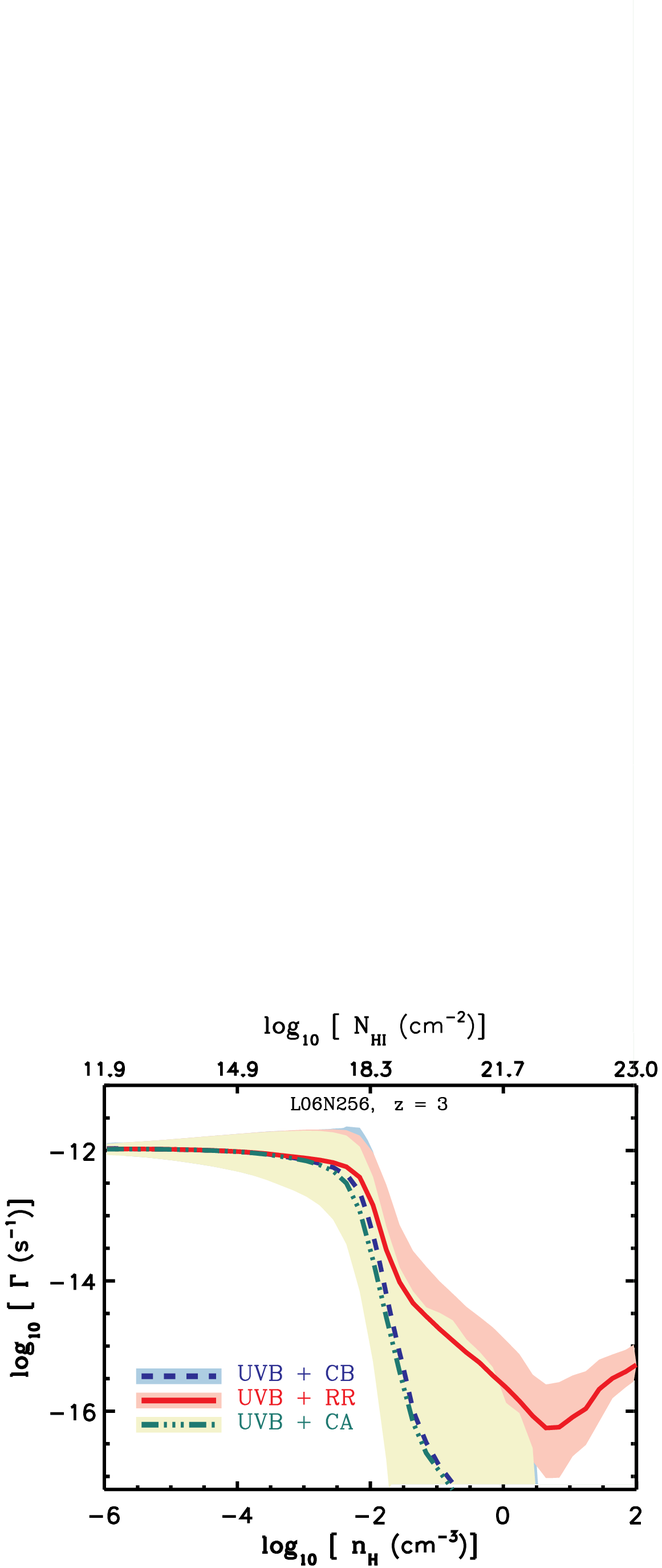}} }
\caption{The hydrogen neutral fractions (\emph{left}) and the UVB photoionization rate profiles (\emph{right}) as a function of density in RT simulations with different models for recombination radiation for the \emph{L06N256} simulation at z = 3. The red dashed curve shows the reference simulation where recombination radiation is modeled self-consistently. The blue solid and green dot-dashed curves show simulations in which recombination radiation is substituted by the use of Case A and Case B recombination rates, respectively. The curves show the medians and the shaded areas around the medians indicate the $15\% - 85\%$ percentiles. HI column densities corresponding to each density are shown along the top x-axes. The effect of recombination radiation on the hydrogen neutral fractions is similar to the use of Case A recombination at low densities (i.e., $\nH \lesssim 10^{-3}\cmcb$) and to the use of Case B recombination at higher densities (i.e., $\nH \gtrsim 10^{-1}\cmcb$). However, recombination radiation can penetrate into the self-shielded regions, an effect that is not captured by the use of Case B recombination. }
\label{fig:Gamma_z3_cases}
\end{figure*}
To study the interplay between different ionizing processes and their effects on the distribution of $\HI$, we compare their ionization rates at different densities. We start the analysis by presenting the results at $z = 3$ and extend it to other redshifts in \S\ref{sec:result-zs}.
\par
The total photoionization rate profiles shown in the right panel of Figure \ref{fig:Gamma_eta_Res} are almost flat at low densities and decrease with increasing density, starting at densities $\nH \sim 10^{-4}\cmcb$. Just below $\nH = 10^{-2}\cmcb$ self-shielding causes a sharp drop, but the fall-off becomes shallower for $\nH > 10^{-2}\cmcb$ and the photoionization rate starts to increase at $\nH > 10\cmcb$. As shown in Figure \ref{fig:Gamma_z3_comp}, the shallower fall-off in the total photoionization rate with increasing density is caused by RR. The increase in the photoionization rate with density at the highest densities on the other hand, is an artifact of the imposed temperature for ISM particles (i.e., $T = 10^4$ K) which produces a rising collisional ionization rate with increasing density. As the comparison between the UVB and RR photoionization profiles shows (see Figure \ref{fig:Gamma_z3_comp}), RR only starts to dominate the total photoionization rate at $\nH > 10^{-2}\cmcb$, where the UVB photoionization rate has dropped by more than one order of magnitude and the gas is no longer highly ionized. RR reduces the total $\HI$ content of high-density gas by $\approx 20\%$. Although ionization rates remain non-negligible at higher densities, they cannot keep the hydrogen highly ionized. For instance at $\nH \sim 1\cmcb$, a photoionization rate of $\Gamma \sim 10^{-14}\ps$ can only ionize the gas by $\lesssim 20\%$.
\par
The shape of the photoionization rate profile produced by diffuse RR can be understood by noting that the production rate of RR increases with the density of ionized gas. At number densities $\nH < 10^{-2}\cmcb$, where the gas is highly ionized, the photoionization rate due to recombination photons is proportional to the density (i.e., $\Gamma_{\rm{RR}} \propto \nH$). At higher densities on the other hand, the gas becomes neutral. As a result, the density of ionized gas decreases with increasing density and the production rate of recombination photons decreases. Therefore, there is a peak in the photoionization rate due to RR around the self-shielding density. At very low densities, the superposition of recombination photons which have escaped from higher densities becomes dominant and the net photoionization rate of recombination photons flattens. Note that our simulations may underestimate this asymptotic rate because our simulation volumes are small compared to the mean free path for ionizing radiation (which is $\sim 100$ Mpc at $z \sim 3$). On the other hand, the neglect of cosmological redshifting for RR will result in overestimation of its photoionization rate on large scales. Recombination photons also leak from lower densities to self-shielded regions, smoothing the transition between highly ionized and highly neutral gas. At high densities, in the absence of the UVB ionizing photons, RR and collisional ionization can boost each other by providing more free electrons and ions. 
\par
In Figure \ref{fig:Gamma_z3_cases} we compare hydrogen neutral fraction and photoionization rate profiles for different assumptions about RR. The hydrogen neutral fraction profile based on a precise RT calculation of RR is close to the Case A result at low densities ($\nH \lesssim 10^{-3} \cmcb$) but converges to the Case B result at high densities ($\nH \gtrsim 10^{-1} \cmcb$). This suggests that the neutral fraction profile, though not the ionization rate, can be modeled by switching from Case A to Case B recombination at $\nH \sim n_{\rm{H, SSh}}$ \cite[e.g.,][]{Altay11, McQuinn11}. 

\subsection{Evolution}
\label{sec:result-zs}
\begin{figure*}
\centerline{\hbox{\includegraphics[width=0.4\textwidth]
             {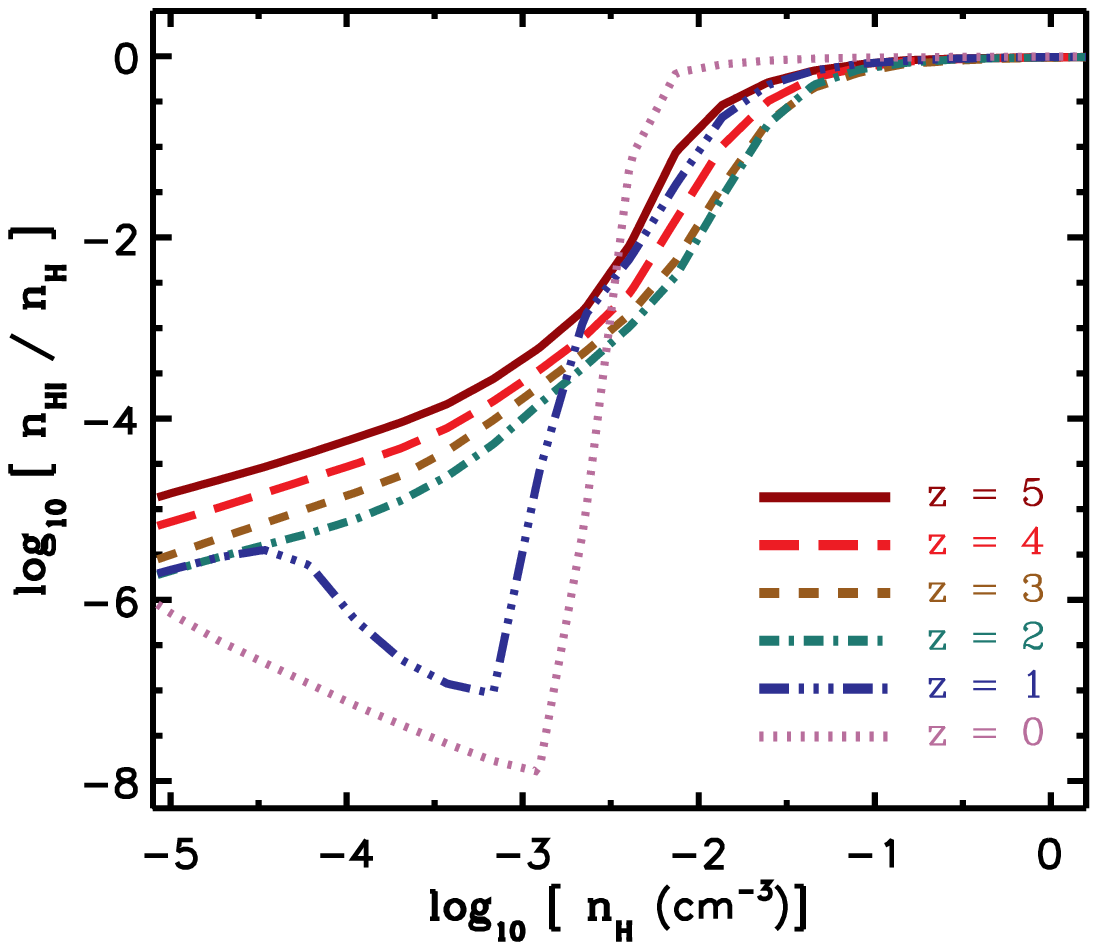}} 
             \hbox{\includegraphics[width=0.4\textwidth]
             {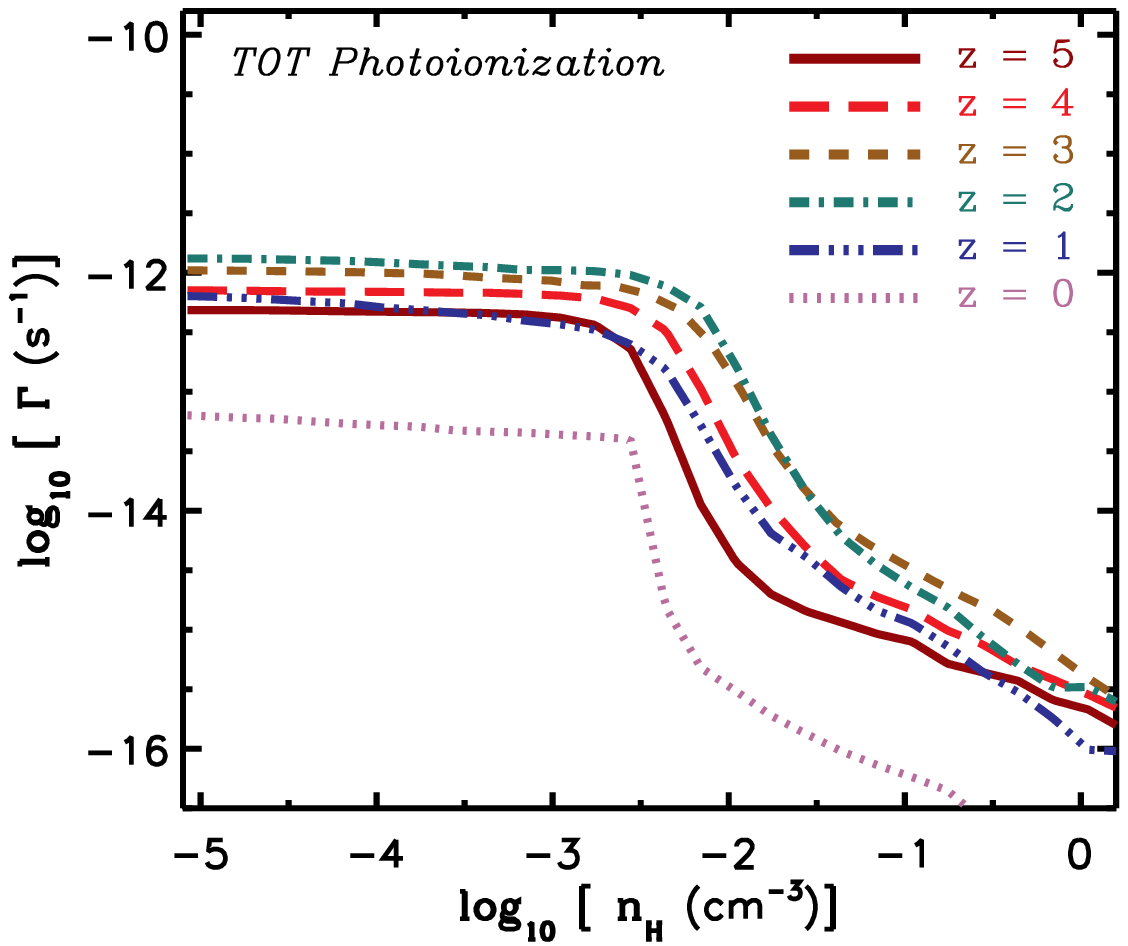}}}
\centerline{\hbox{\includegraphics[width=0.4\textwidth]
             {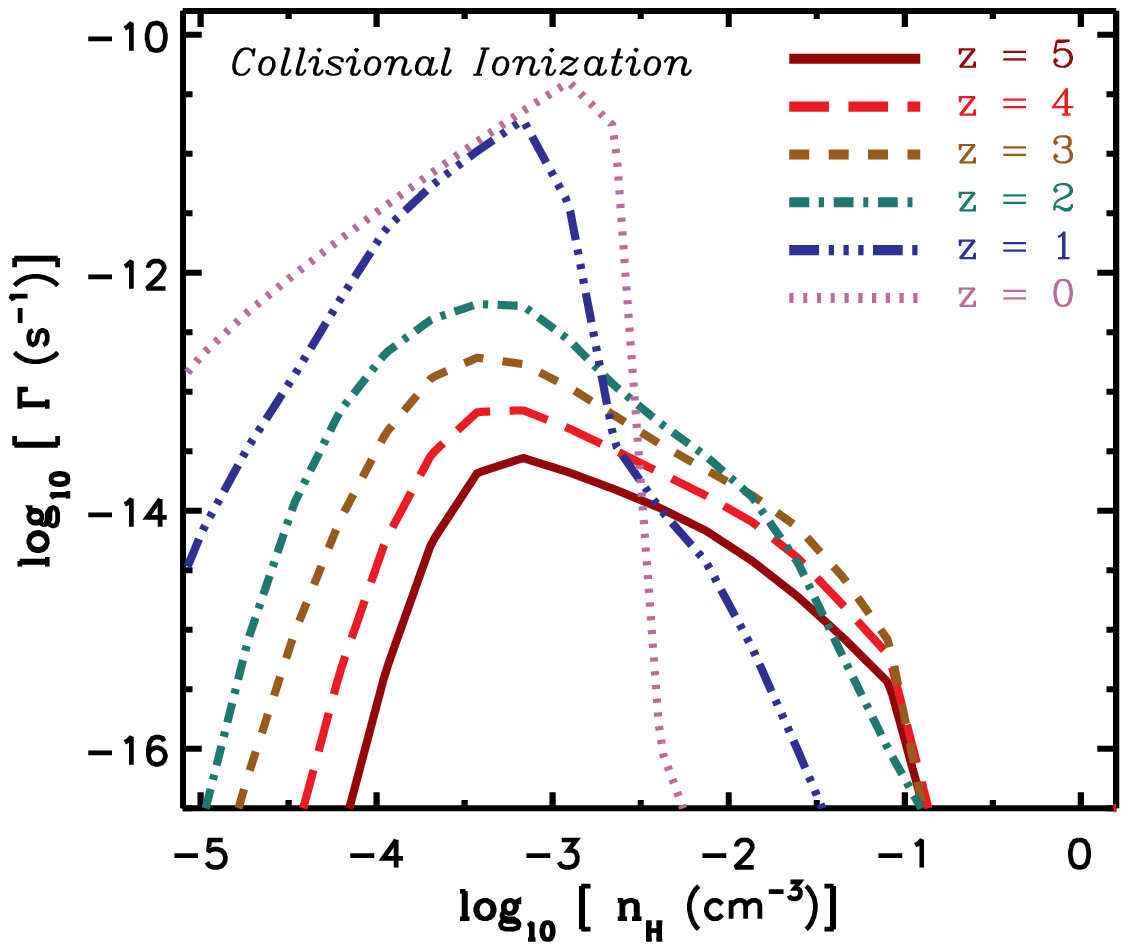}} 
             \hbox{\includegraphics[width=0.4\textwidth]
             {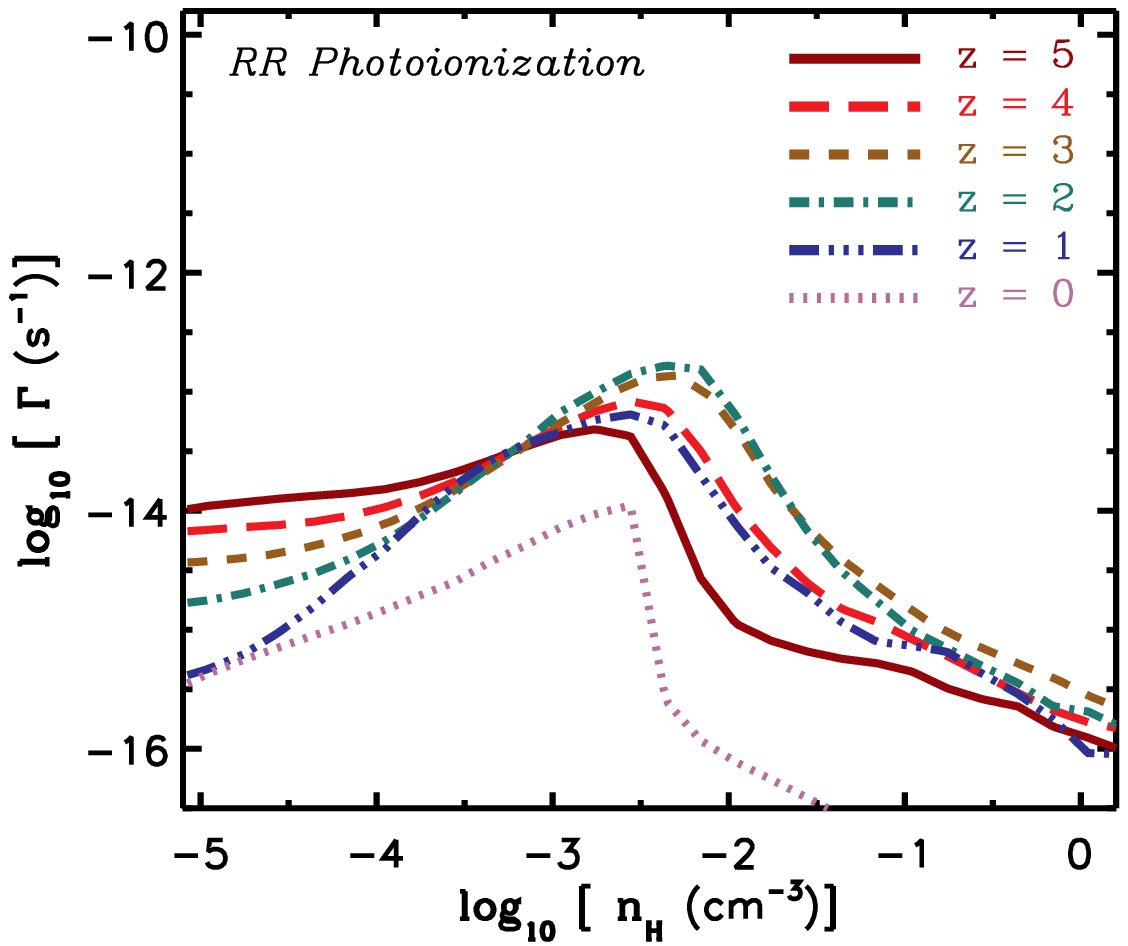}}}
\caption{Evolution of the hydrogen neutral fraction profile (\emph{top-left}) and various ionization rates as a function of density. \emph{Top-right}, \emph{bottom-left} and \emph{bottom-right} panels show, respectively, the total photoionization rates, collisional ionization rates and RR photoionization rates. All the $\HI$ fractions and collisional ionization rates which are sensitive to both collisional ionization and photoionization are taken from the \emph{L50N512-W3} simulation. Photoionization rates at $z \geq 2$ are based on the \emph{L06N128} simulation. At lower redshifts (i.e., $z = 0$ and 1), where the box size become important because of the collisional ionization and its effect on changing the self-shielding, we used a representative sub-volume of the \emph{L50N512-W3} simulation to calculate the photoionization rate profile. with density While the overall shape of the UVB photoionization rate profile is similar at different redshifts, the collisional ionization becomes increasingly stronger at lower redshifts and  strongly reduces the hydrogen neutral fractions at densities $\nH \lesssim 10^{-3}\cmcb$. }
\label{fig:eta_gamma_compz}
\end{figure*}
The general trends in the profile of the photoionization rates with density and their influences on the distribution of $\HI$ are not very sensitive to redshift. However, as shown in table \ref{table:UVBz}, the intensity and hardness of the UVB radiation change with redshift which, in turn, changes the self-shielding density. Moreover, as the Universe expands, the average density of absorbers decreases and their distributions evolve. The larger structures that form at lower redshifts drastically change the temperature structure of the gas at low and intermediate densities where collisional ionization becomes the dominant process. In the top-left panel of Figure \ref{fig:eta_gamma_compz}, the evolution of the hydrogen neutral fraction is illustrated for the \emph{L50N512-W3} simulation. As discussed in $\S$\ref{sec:Photoionization-density-fit} and Appendix \ref{ap:RT-fit-tests}, since the photoionization rate profiles are converged with box size and resolution, we apply the profiles derived from a RT simulation of a smaller box, or a subset of the big box at lower redshifts\footnote{Strong collisional ionization at low redshifts can change the self-shielding. Therefore, for our RT simulation at low redshifts (i.e., $z\lesssim 1$) we used a representative sub-volume of the \emph{L50N512-W3} simulation sufficiently large for the collisional ionization rates to be converged.}, to calculate the neutral fractions in this big box. Figure \ref{fig:eta_gamma_compz} shows that the neutral fraction profiles are similar in shape at high redshifts but that at $z \leq 1$ the profiles are largely different, particularly at low hydrogen number densities, due to the evolving collisional ionization rates.
\par
The evolution of the collisional ionization rate profiles is shown in the bottom-left panel of Figure \ref{fig:eta_gamma_compz}.  At $z \geq 2$ and for $\nH < 10^{-2}\cmcb$, the collisional ionization rate is not high enough to compete with the UVB photoionization rate. At lower redshifts and for number densities $\nH \lesssim 3\times10^{-3}\cmcb$, on the other hand, collisional ionization dominates\footnote{We note that at redshift $z \lesssim 1$ box sizes $L_{\rm{Box}} \gtrsim 25$ comoving $\Mpch$ are required to fully capture the large-scale accretion shocks and to produce converged collisional ionization rates.}. Indeed, the median collisional ionization rates are more than $100$ times higher than the UVB photoionization rate at densities around the expected self-shielding thresholds. Collisional ionization therefore helps the UVB ionizing photons to penetrate to higher densities without being significantly absorbed. As a result, self-shielding starts at densities higher than expected from equation \eqref{eq:densitySSH}. The signature of collisional ionization on the hydrogen neutral fraction is more dramatic at low densities and partly compensates for the lower UVB intensity at z = 0. This results in a flattening of $\fNHI$ at column densities $\NHI \lesssim 10^{16}\cmsq$ as shown in the left panel of Figure \ref{fig:column_obs_box}.
\par
As mentioned above, at low redshifts (e.g., $z =0$) the collisional ionization rate peaks at densities higher than the expected self-shielding threshold against the UVB. As a result, the total photoionization rate falls off rapidly together with the drop in the collisional ionization rate. Therefore, the drop in the hydrogen ionized fraction, and hence the resulting free electron density, is much sharper at lower redshifts. This causes a steeper high-density fall-off in the collisional ionization rate as shown in the bottom-left panel of Figure \ref{fig:eta_gamma_compz}.
\par
The differences between the total photoionization rates at different redshifts shown in the top-right panel of Figure \ref{fig:eta_gamma_compz}, are caused by the evolution of the UVB intensity and its hardness, which affects the self-shielding density thresholds (see equation \ref{eq:densitySSH}). On the other hand, as we showed in the previous section, the peak of the photoionization rate produced by RR tracks the self-shielding density. As a result the peak of the RR photoionization rate also changes with redshift as illustrated in the bottom-right panel of Figure \ref{fig:eta_gamma_compz}). 
\par
\begin{figure}
\centerline{\hbox{\includegraphics[width=0.45\textwidth]
             {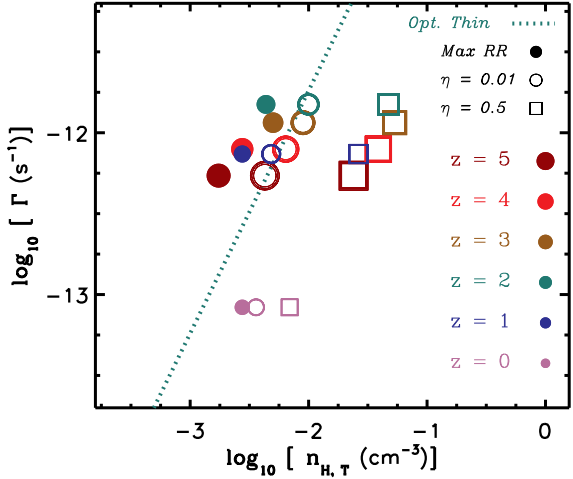}} }
\caption{The photoionization from recombination radiation peaks at the expected self-shielding density and smooths the transition between highly ionized gas and self-shielded gas. This is illustrated by showing different characteristic densities for various redshifts. Filled circles show the density at which recombination radiation peaks, while open circles and squares show the $\HI$ number density corresponding to hydrogen neutral fractions of $10^{-2}$ and $0.5$, respectively. The self-shielding density threshold for a given photoionization rate expected from the Jeans scaling argument (equation \ref{eq:densitySSH}) is also indicated by the green dotted line. The Jeans scaling argument works well, except at $z = 0$ when collisional ionization is important.}
\label{fig:self_shielding}
\end{figure}

The filled circles in Figure \ref{fig:self_shielding} indicate the UVB photoionization rate versus the number density at which the RR photoionization rate peaks. The self-shielding density expected from the Jeans scaling argument (equation \ref{eq:densitySSH}) is also shown (green dotted line). The peaks in the RR photoionization rate in RT simulations follow this expected scaling for $z \geq 1$. However, the $z = 0$ result deviates from this trend since collisional ionization affects the self-shielding density threshold, a factor that is not captured by equation \eqref{eq:densitySSH}.
\par
As a result of the RR photoionization rate peaking around the self-shielding density threshold, the transition between highly ionized and nearly neutral gas becomes more extended at all redshifts. To illustrate the smoothness of this transition, the densities at which the median hydrogen neutral fractions are $10^{-2}$ and $0.5$ are shown in Figure  \ref{fig:self_shielding} with open circles and squares,  respectively. The densities at which the hydrogen neutral fraction is $10^{-2}$ are slightly higher than the densities at which the RR photoionization rate peaks (filled circles). The evolution agrees with the trend expected from the Jeans scaling argument and the self-shielding density. The exception is again $z = 0$, where the large collisional ionization rate at number densities $\nH \sim 10^{-3}-10^{-2}\cmcb$ shifts the transition to neutral fraction of 0.5 to densities that are $\approx 1$ dex higher. However, the relation between the photoionization rate and the density still follows the slope expected from equation \eqref{eq:densitySSH}. 
\par
It is interesting to note that the UVB spectral shape at $z =4$ is slightly harder than at $z = 1$ while the UVB intensities at these two redshifts are similar. This results in a deeper penetration of ionizing photons at $z = 4$. Consequently, the densities corresponding to the indicated neutral fractions (i.e., $10^{-2}$ and $0.5$) at $z = 1$ are lower than their counterparts at $z = 4$.
\section{Conclusions}
\label{sec:conclusions}
We combined a set of cosmological hydrodynamical simulations with an accurate RT simulation of the UVB radiation to compute the $\HI$ column density distribution function and its evolution. We ignored the effect of local sources of ionizing radiation, but we did include a self-consistent treatment of recombination radiation. 
\par
Our RT results for the distribution of photoionization rates at different densities are converged with respect to the simulation box size and resolution. Therefore, the resulting photoionization rate can be expressed as a function of the hydrogen density and the UVB. We provided a fit for the median total photoionization rate as a function of density that can be used with any desired UVB model to take into account the effect of HI self-shielding in cosmological simulations without the need to perform RT.
\par
The CDDF, $\fNHI$, predicted by our RT simulations is in excellent agreement with observational constraints at all redshifts ($z = 0 - 5$) and reproduces the slopes of the observed $\fNHI$ function for a wide range of HI column densities. At low HI column densities, the CDDF is a steep function which decreases with increasing $\NHI$ before it flattens at $\NHI \gtrsim 10^{18} \cmsq$ due to self-shielding. At $\NHI \gtrsim 10^{21}\cmcb$ on the other hand, $\fNHI$ is determined mainly by the intrinsic distribution of total hydrogen and the $\rm{H_2}$ fraction. 
\par
We showed that the $\NHI-\nH$ relationship can be explained by a simple Jeans scaling. This argument assumes $\HI$ absorbers to be self-gravitating systems close to local hydrostatic equilibrium \citep{Schaye01} and to be either neutral or in photoionization equilibrium in the presence of an ionizing radiation field. However, at $z =0$ the analytic treatment underestimates the self-shielding density threshold due to its neglect of collisional ionization.
\par
The high $\HI$ column density end of the predicted $\fNHI$ evolves only weakly from $z = 5$ to $z = 0$, consistent with observations. In the Lyman limit range of the distribution function, the slope of $\fNHI$ remains the same at all redshifts. However,  at $z > 3$ the number of absorbers increases with redshift as the Universe becomes denser while the UVB intensity remains similar. At lower redshifts, on the other hand, the combination of a decreasing UVB intensity and the expansion of the Universe results in a non-evolving $\fNHI$. In contrast, the number of absorbers with lower $\HI$ column densities (i.e., the Ly$\alpha$ forest) decreases significantly from $z\sim3$. We showed that this results in part from the stronger collisional ionization at redshifts $z \lesssim 1$, which compensates for the lower intensity of the UVB. The increasing importance of collisional ionization is due to the rise in the fraction of hot gas due to shock-heating associated with the formation of structure.
\par
The inclusion of diffuse recombination radiation smooths the transition between optically thin and thick gas. Consequently, the transition to highly neutral gas is not as sharp as what has been assumed in some previous works \citep[e.g.,][]{Nagamine10,Yajima11,Goerdt12}. For instance, the difference in the gas density at which hydrogen is highly ionized (i.e., $\nHI/\nH \lesssim 0.01$) and the density at which gas is highly neutral (i.e., $\nHI/\nH \gtrsim 0.5$) is more than one order of magnitude (see Figure \ref{fig:self_shielding}). As a result, assuming a sharp self-shielding density threshold at the density for which the optical depth of ionizing photons is $\sim 1$, overestimates the resulting neutral hydrogen mass by a factor of a few.
\par
Our simulations adopted some commonly used approximations (e.g., neglecting helium RT effects, using a gray approximation in order to mimic the UVB spectra, neglecting absorption by dust and local sources of ionizing radiation). Our tests show that most of those approximations have negligible effects on our results. But there are some assumptions which require further investigation. For instance, the presence of young stars in high-density regions could change the HI CDDF, especially at high HI column densities through feedback and emission of ionizing photons. Indeed, we will show in Rahmati et al. (in prep.) that for very high column densities the ionizing radiation from young stars can reduce the $\fNHI$ by 0.5-1 dex. 

\section*{Acknowledgments}
We thank the anonymous referee for a helpful report. We thank Garbriel Altay for providing us with his simulation results and a compilation of the observed HI CDDF. We also would like to thank Kristian Finlator, J. Xavier Prochaska, Tom Theuns and all the members of the OWLS team for valuable discussions and Marcel Haas, Joakim Rosdahl, Maryam Shirazi and Freeke van de Voort  for helpful comments on an earlier version of the paper. The simulations presented here were run on the Cosmology Machine at the Institute for Computational Cosmology in Durham (which is part of the DiRAC Facility jointly funded by STFC, the Large Facilities Capital Fund of BIS, and Durham University) as part of the Virgo Consortium research programme and on Stella, the LOFAR BlueGene/L system in Groningen. This work was sponsored by the National Computing Facilities Foundation (NCF) for the use of supercomputer facilities, with financial support from the Netherlands Organization for Scientific Research (NWO), also through a VIDI grant and an NWO open competition grant. We also benefited from funding from NOVA, from the European Research Council under the European UnionÕs Seventh Framework Programme (FP7/2007-2013) / ERC Grant agreement 278594-GasAroundGalaxies and from the Marie Curie Training Network CosmoComp (PITN-GA-2009-238356). AHP receives funding from the European Union's Seventh Framework Programme (FP7/2007-2013)  under grant agreement number 301096-proFeSsOR.

\appendix
\section{Photoionization rate as a function of density}
\subsection{Replacing the RT simulations with a fitting function}
\label{ap:RT-fit-tests}
In $\S$\ref{sec:Photoionization-density-fit} we demonstrated that the median of the simulated relation between the total photoionization rate, $\Gamma_{\rm{Phot}}$, and density is converged with respect to resolution and box size. We used this result and provided fits to the median of this relation. We have exploited these fits to compute the neutral hydrogen fraction in cosmological simulations under the assumption of ionization equilibrium (see Appendix \ref{ap:equilib_neut}), without performing the computationally demanding RT. In this section, we discuss the accuracy of these fits.
\begin{figure*}
\centerline{\hbox{\includegraphics[width=0.5\textwidth]
              {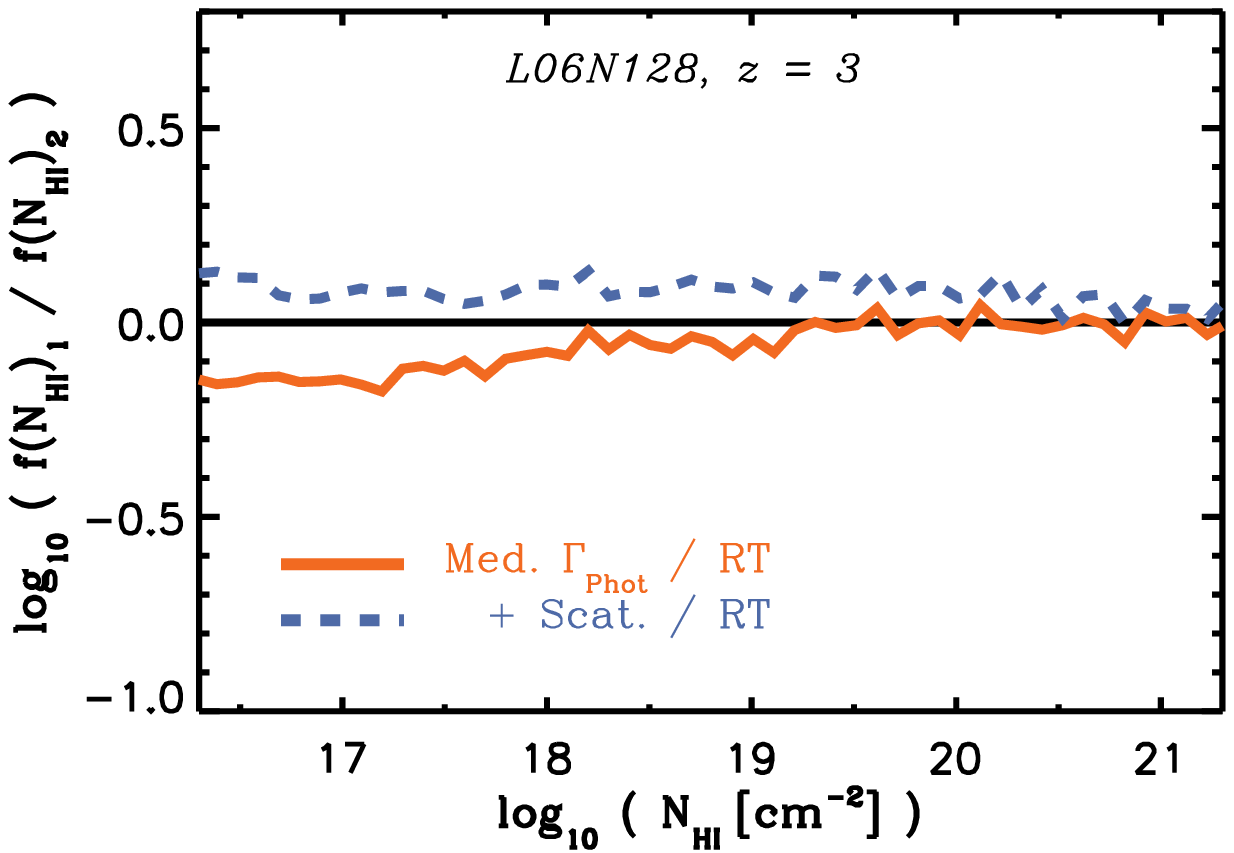}}
             \hbox{\includegraphics[width=0.5\textwidth]
             {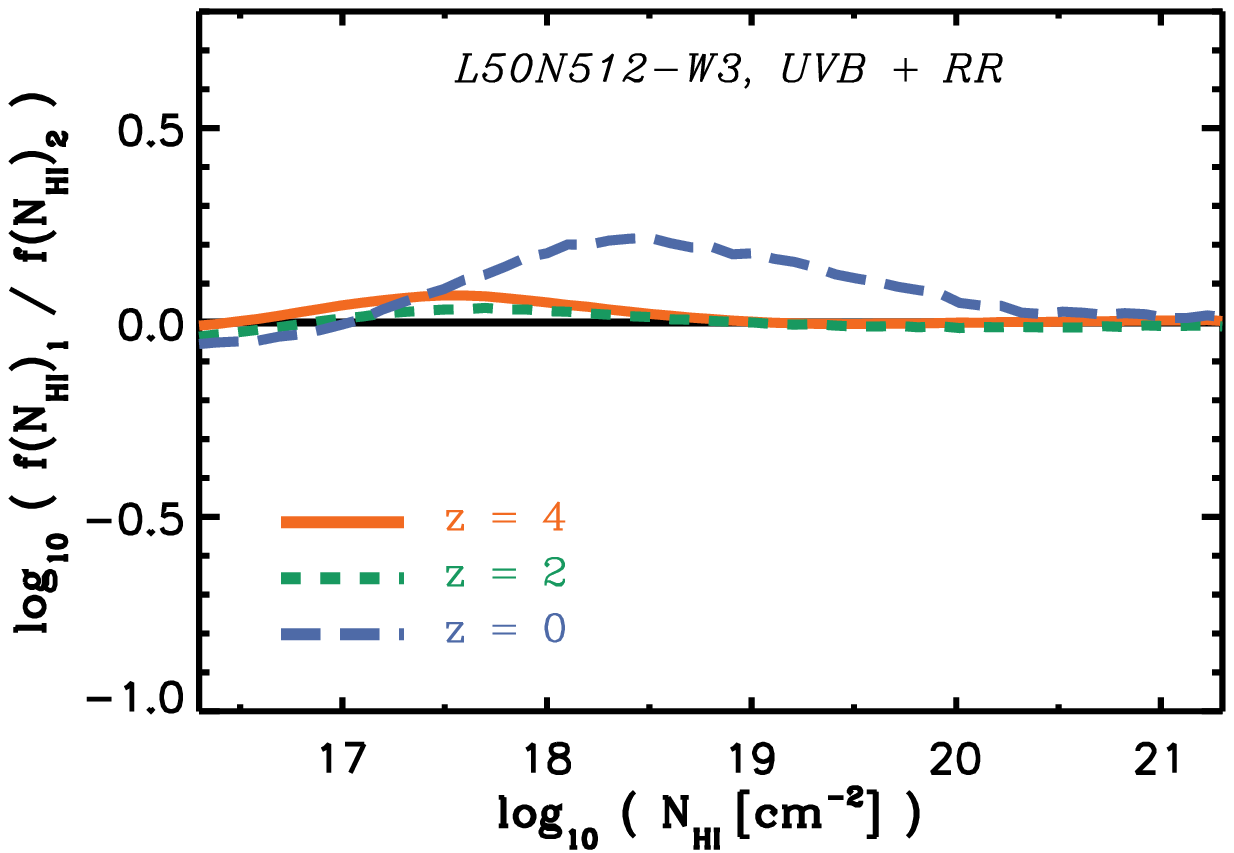}}}
\caption{\emph{Left:} The ratio between the HI CDDF calculated using the RT based $\Gamma_{\rm{Phot}}-\nH$ relationship and the actual RT results for the \emph{L06N128} simulation in the presence of the UVB and diffuse recombination radiation, at $z =3$. The orange solid line shows the result of using the median photoionization rate-density profile predicted by the RT simulation. The blue dashed curve shows the result of including the scatter around the median in the calculations. \emph{Right:} HI CDDFs calculated using the $\Gamma_{\rm{Phot}}-\nH$ fitting function (i.e., equation \ref{eq:Gamma-fit}) are compared to the HI CDDFs for which the actual $\Gamma_{\rm{Phot}}$-density relation from the RT simulations are used. Blue and green curves are for $z=0$ and $z =2$ respectively and the red curve is for $z =4$. The difference between the RT result and the result of using the fitting function at $z = 0$ is due to the importance of collisional ionization at $z = 0$. To capture this effect and to reproduce the RT results at $z = 0$, we advise using the best-fit parameters shown in Table \ref{table:best-fitz0}. All the CDDFs are for the \emph{L50N512-W3} simulation and in the presence of the UVB and diffuse recombination radiation.}
\label{fig:Cfit-test}
\end{figure*}
\par
The left panel of Figure~\ref{fig:Cfit-test} shows that using the median photoionization rates produces an HI CDDF in very good agreement with the HI CDDF obtained from the corresponding RT simulation (orange solid curve) at $\NHI \gtrsim 10^{18}\cmsq$. However, there is a small systematic difference at lower column densities. One may think that this small difference is caused by the loss of information contained in the scatter in the photoionization rates at fixed density. We tested this hypothesis by including a log-normal random scatter around the median photoionization rate consistent with the scatter exhibited by the RT result. However, after accounting for the random scatter, the $\fNHI$ is slightly overproduced compared to the full RT result at nearly all $\HI$ column densities. 
\par
\begin{figure*}
\centerline{\hbox{\includegraphics[width=0.5\textwidth]
              {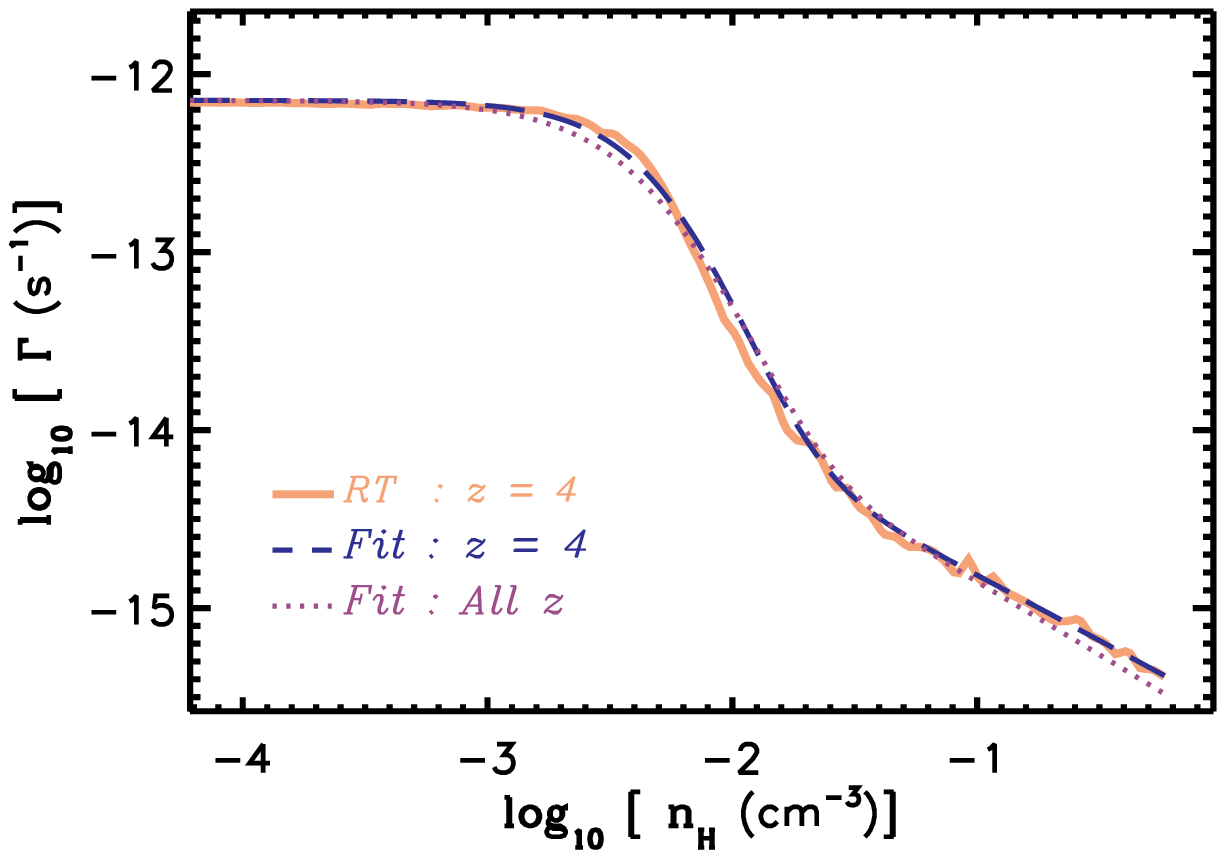}}
             \hbox{\includegraphics[width=0.5\textwidth]
             {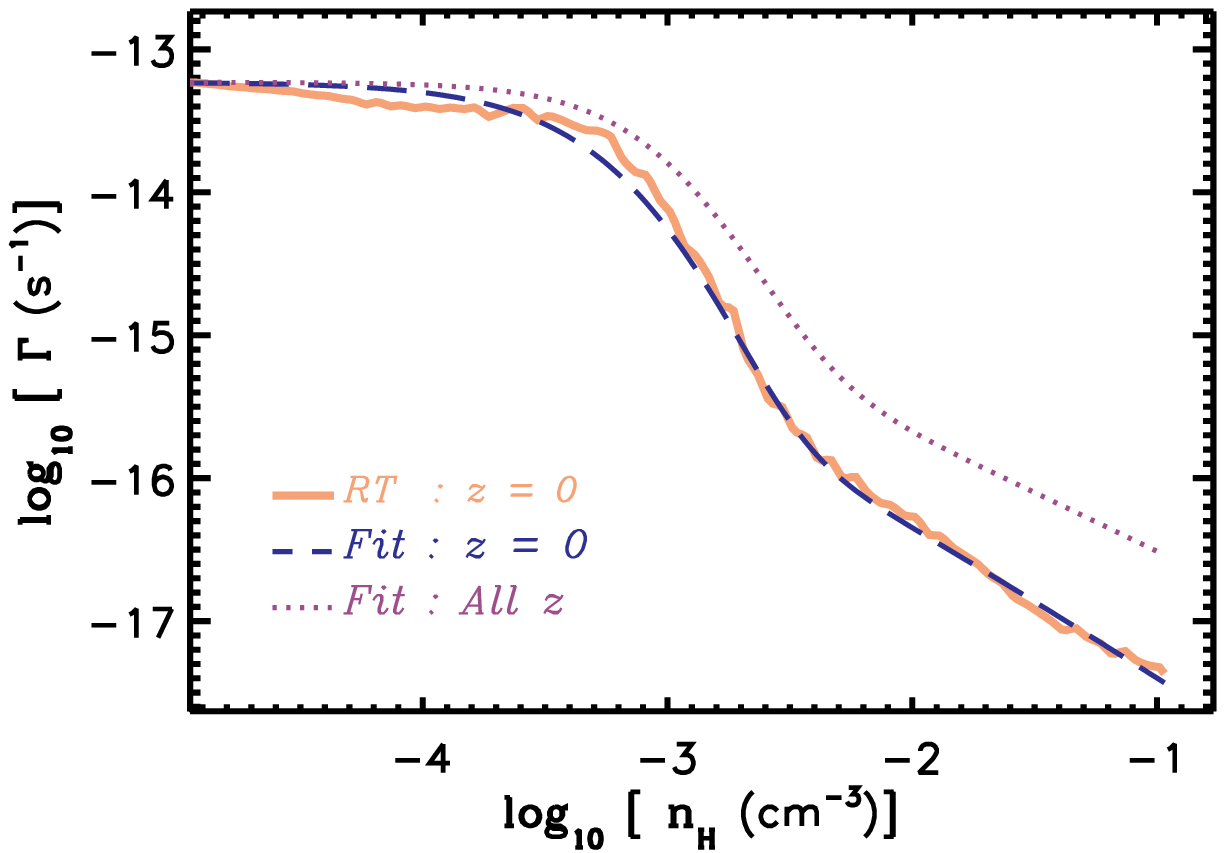}}}
\caption{Comparisons between the total photoionization rates as a function of density in the \emph{L06N128} simulation. Photoionization rates based on the RT simulations and best-fit functions at $z = 4$ and $ z = 0$ are shown in the \emph{left} and \emph{right} panels, respectively. In each panel, the RT result is shown with the orange solid curve. The best fit to the RT result at a given redshift (equation \ref{eq:Gamma-fit0} and Table \ref{table:best-fit}) is shown with the blue dashed curve and the best fit to the RT results at $z = 1 - 5$ (equation \ref{eq:Gamma-fit}) is shown with the purple dotted curve. As shown in the \emph{right} panel, because of the impact of collisional ionization on self-shielding, the low redshift photoionization curve (the blue dashed curve) deviates from the best fit to the results at higher redshifts (the purple dotted curve). To resolve this issue and to capture the impact of collisional ionization, we advise using the best-fit parameters shown in Table \ref{table:best-fitz0} for $z = 0$.}
\label{fig:best-fit}
\end{figure*}

We exploit the insensitivity of the shape of the $\Gamma_{\rm{Phot}}$-density relation to the redshift, and propose the following fit to the photoionization rate, $\Gamma_{\rm{Phot}}$,
\begin{equation}
\frac{\Gamma_{\rm{Phot}}} {\Gamma_{\rm{UVB}}} = (1-f) \left[1 + \left(\frac{\nH}{n_0}\right)^{\beta}  \right]^{\alpha_1} +f \left[1 + \frac{\nH}{n_0} \right]^{\alpha_2},
\label{eq:Gamma-fit0}
\end{equation}
where $\Gamma_{\rm{UVB}}$ is the photoionization rate due to the ionizing background, and $n_0$, $\alpha_1$, $\alpha_2$, and $\beta$ are parameters of the fit. The best-fit values of these parameters are listed in Table \ref{table:best-fit} and the photoionization rate-density relations they produce are compared with the RT simulations at redshifts $z = 0$ and $z = 4$ in Figure~\ref{fig:best-fit}. At all redshifts, the best-fit value of $n_0$ is almost identical to the self-shielding density threshold, $n_{\rm{H, SSh}}$, defined in equation \eqref{eq:densitySSH}, and the characteristic slopes of the photoionization rate-density relation are similar. This suggest that one can find a single set of best-fit values to reproduce the RT results at $z \gtrsim 1$. The corresponding best-fit parameter values are (see also equation 
\ref{eq:Gamma-fit}) $\alpha_1 = -2.28 \pm 0.31$, $\alpha_2 = -0.84 \pm 0.11$, $n_0 = (1.003 \pm 0.005)\times n_{\rm{H, SSh}}$, $\beta = 1.64 \pm 0.19$ and $f = 0.02 \pm 0.0089$.
\par
In the right panel of Figure~\ref{fig:Cfit-test}, the ratio between the HI CDDF calculated using the fitting function presented in equation \eqref{eq:Gamma-fit} and the RT based HI CDDF (i.e., calculated using the median of the photoionization rate-density relation in the RT simulations) is shown for the \emph{L50N512-W3} and at $z = 0,~2$ and 4. This illustrates that the fitting function reproduces the RT results accurately, except at $z = 0$. As explained in $\S$\ref{sec:results}, this is expected since at low redshifts collisional ionization affects the self-shielding and the resulting photoionization rate-density profile. However, a separate fit can be obtained using converged RT results at $z = 0$. The parameters that define such a fit are shown in Table \ref{table:best-fitz0}.
\begin{table}
\caption{The best-fit parameters for equation \eqref{eq:Gamma-fit0} at different redshifts based on RT results in the \emph{L06N128} simulation.}
\begin{tabular}{c c c c c c}
\hline
Redshift & $\log{[n_0]}~(\cmcb)$ & $\alpha_1$ &$\alpha_2$& $\beta$ &$1-f$\\
\hline
\hline
z = 1-5 & $\log{[n_{\rm{H, SSh}}]}$ & -2.28 & -0.84 & 1.64 &  0.98\\
 \hline
$z =0$ & -2.94 &  -3.98 & -1.09 & 1.29 &  0.99\\
$z =1$ & -2.29 &  -2.94 & -0.90 & 1.21 & $ 0.97$\\
$z =2$ & -2.06 &  -2.22 & -1.09 & 1.75 & $ 0.97$\\
$z =3$ & -2.13 &  -1.99 & -0.88 & 1.72 & $ 0.96$\\
$z =4$ & -2.23 &  -2.05 & -0.75 & 1.93 & $ 0.98$\\
$z =5$ & -2.35 &  -2.63 & -0.57 & 1.77 & $ 0.99$\\
\hline
\end{tabular}
\label{table:best-fit}
\end{table}
\begin{table}
\caption{The best-fit parameters for equation \eqref{eq:Gamma-fit0} at $z = 0$ based on RT results for the \emph{L50N512} simulation. To capture the impact of collisional ionization on the self-shielding, one needs to use large cosmological simulations. The simulation with a box size of $50~\Mpch$ results in converged collisional ionizations.}
\begin{tabular}{c c c c c c}
\hline
Redshift & $\log{[n_0]}~(\cmcb)$ & $\alpha_1$ &$\alpha_2$& $\beta$ &$1-f$\\
\hline
\hline
$z =0$ & -2.56 &  -1.86 & -0.51 & 2.83 &  0.99\\
\hline
\end{tabular}
\label{table:best-fitz0}
\end{table}
\subsection{The equilibrium hydrogen neutral fraction}
\label{ap:equilib_neut}
In this section we explain how to derive the neutral fraction in ionization equilibrium. Equating the total number of ionizations per unit time per unit volume with the total number of recombinations per unit time per unit volume, we obtain
\begin{equation}
n_{\rm{HI}}~\Gamma_{\rm{TOT}} = \alpha_{\rm{A}}~ n_{\rm{e}}~ n_{\rm{HII}},
\label{eq:equilib_rec_ion}
\end{equation}
where $\nHI$, $\nel$ and $\nHII$ are the number densities of neutral hydrogen atoms, free electrons and protons, respectively. $\Gamma_{\rm{TOT}}$ is the total ionization rate per neutral hydrogen atom and $\alpha_{\rm{A}}$ is the Case A recombination rate\footnote{The use of Case B is more appropriate for $\nH > n_{\rm{H, SSh}}$. However, we assume the photoionization due to RR is included in $\Gamma_{\rm{TOT}}$, e.g., by using the best-fit function that is presented in equation \ref{eq:Gamma-fit}. Therefore, Case A recombination should be adopted even at high densities.} for which we use the fitting function given by \citet{Hui97}:
\begin{equation}
 \alpha_{\rm{A}} = 1.269\times 10^{-13}~ \frac{\lambda^{1.503}}{\left(1 + \left(\lambda/0.522\right)^{0.47}\right)^{1.923}}~\cmc \ps,
\label{eq:equilib_rec_ion}
\end{equation}
where $\lambda = 315614/T$.
\par
Defining the hydrogen neutral fraction as the ratio between the number densities of neutral hydrogen and total hydrogen, $\eta = \nHI / \nH$, and ignoring helium (which is an excellent approximation, see Appendix \ref{sec:Helium}), we can rewrite equation \eqref{eq:equilib_rec_ion} as:
\begin{equation}
\eta~\Gamma_{\rm{TOT}} = \alpha_{\rm{A}}~ (1 - \eta)^2~ \nH.
\label{eq:equilib_rec_ionI}
\end{equation}
Furthermore, we can assume that the total ionization rate, $\Gamma_{\rm{TOT}}$, consists of two components: the total photoionization rate, $\Gamma_{\rm{Phot}}$, and the collisional ionization rate, $\Gamma_{\rm{Col}}$:
\begin{equation}
\Gamma_{\rm{TOT}} = \Gamma_{\rm{Phot}} + \Gamma_{\rm{Col}},
\label{eq:equilib_rec_ionII}
\end{equation}
where $\Gamma_{\rm{Col}}=\Lambda_{\rm{T}}~ (1-\eta)~\nH $. The photoionization rate can be expressed as a function of density using equation \eqref{eq:Gamma-fit}. For $\Lambda_{\rm{T}}$, which depends only on temperature, we use a relation given in \citet{Theuns98}:
\begin{equation}
\Lambda_{\rm{T}} = 1.17 \times 10^{-10}~\frac{T^{1/2} \exp(-157809/T)} {1 + \sqrt{T/ 10^5} } ~\cmc \ps.
\label{eq:gamma_colT}
\end{equation}
\par
We can now rearrange equation \eqref{eq:equilib_rec_ionI} as a quadratic equation:
\begin{equation}
A~ \eta^2 - B~ \eta + C = 0,
\label{eq:equilib_quad}
\end{equation}
with $A = \alpha_{\rm{A}} + \Lambda_{\rm{T}} $, $B = 2 \alpha_{\rm{A}} + \frac{\Gamma_{\rm{Phot}}}{\nH} +  \Lambda_{\rm{T}} $ and $C =  \alpha_{\rm{A}}$ which gives:
\begin{equation}
 \eta = \frac{B - \sqrt{B^2-4AC}}{2A}.
\label{eq:equilib_eta}
\end{equation}
\par
Using the last equation one can calculate the equilibrium hydrogen neutral fraction for a given $\nH$ and temperature.
\section{The effects of box size, cosmological parameters and resolution on the HI CDDF}
\label{ap:res-box-cos-tests}
\begin{figure*}
\centerline{\hbox{\includegraphics[width=0.45\textwidth]
             {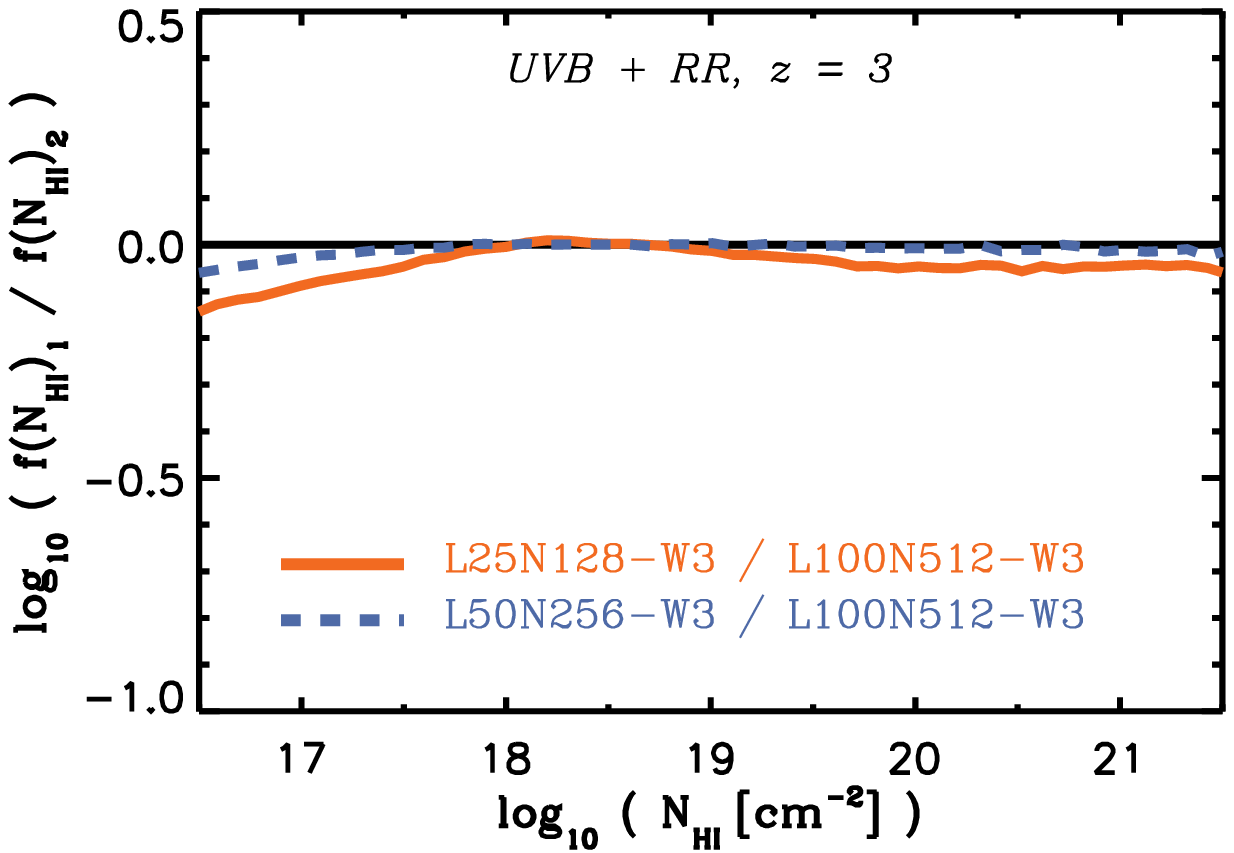}} 
             \hbox{\includegraphics[width=0.45\textwidth]
             {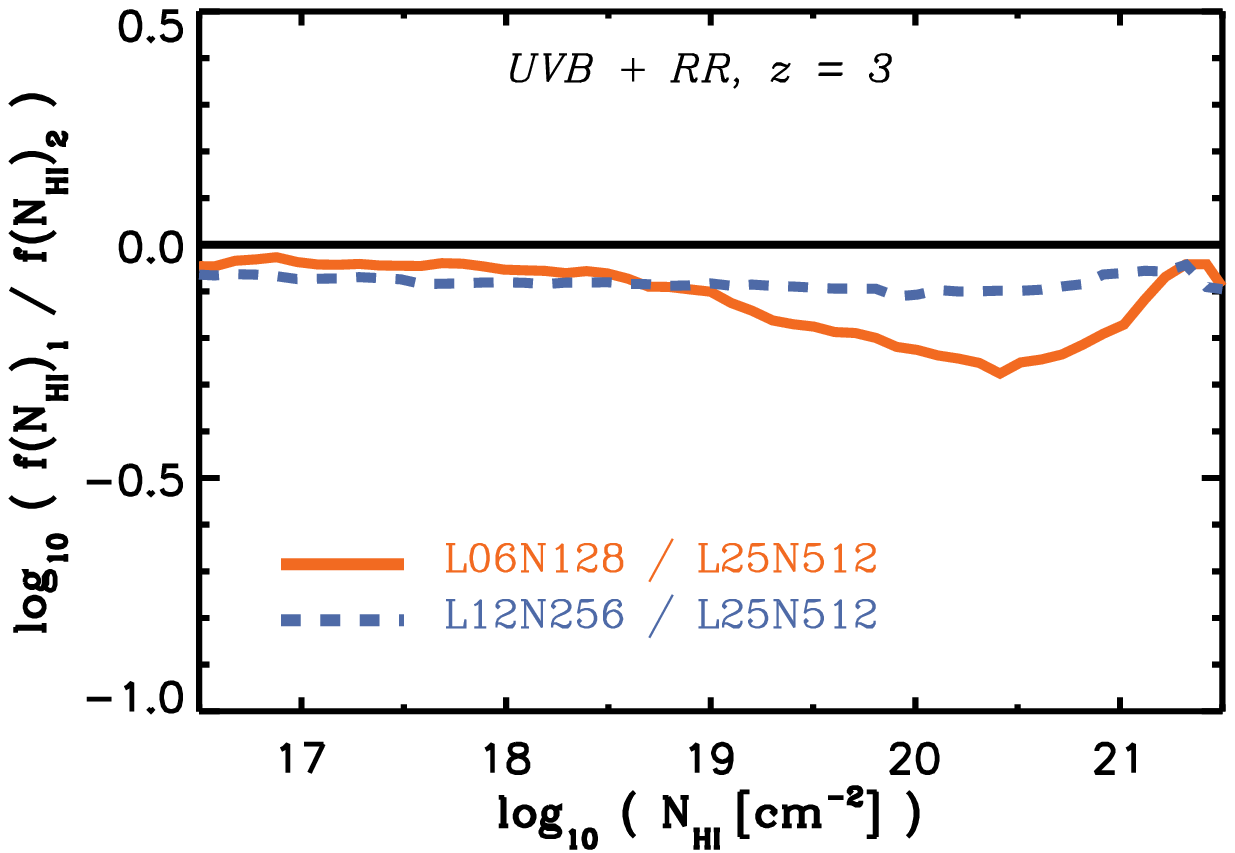}} }
\centerline{\hbox{\includegraphics[width=0.45\textwidth]
             {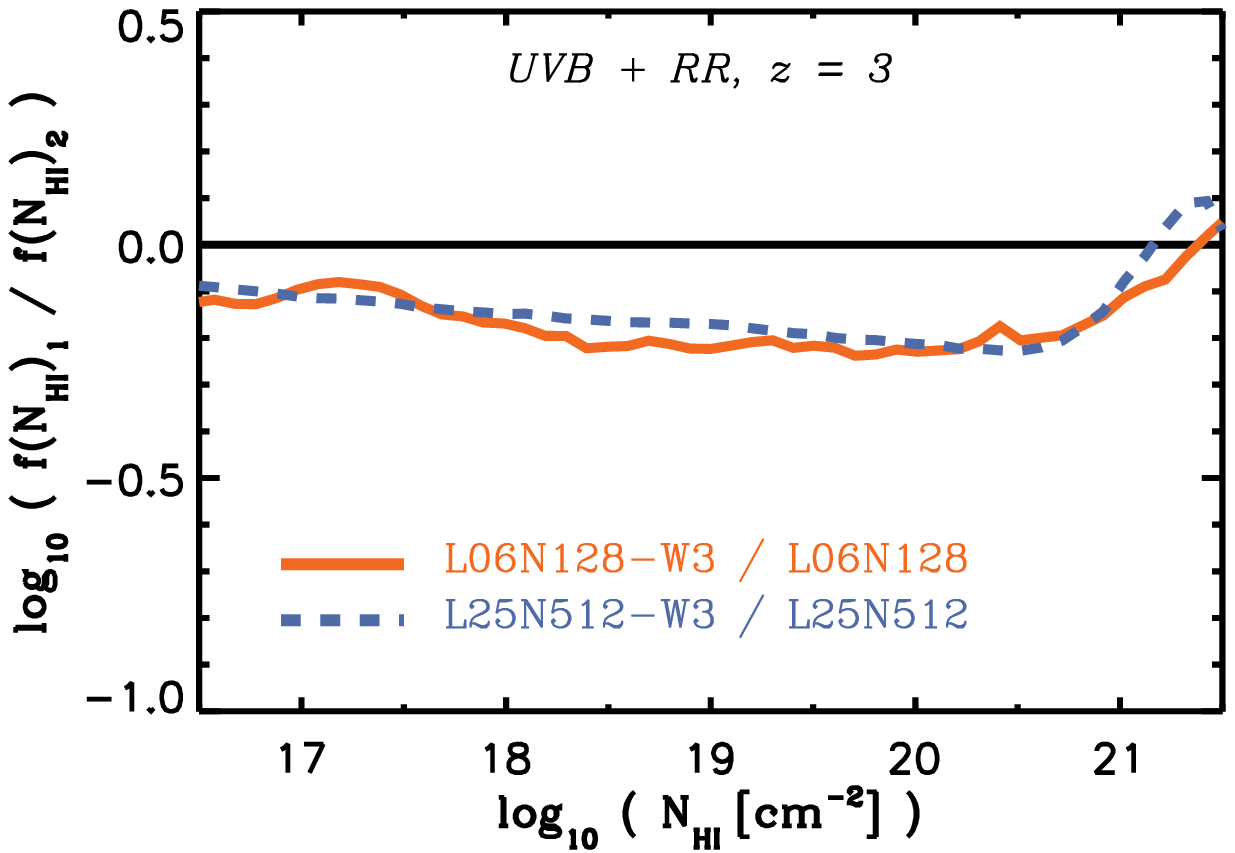}} 
             \hbox{\includegraphics[width=0.45\textwidth]
             {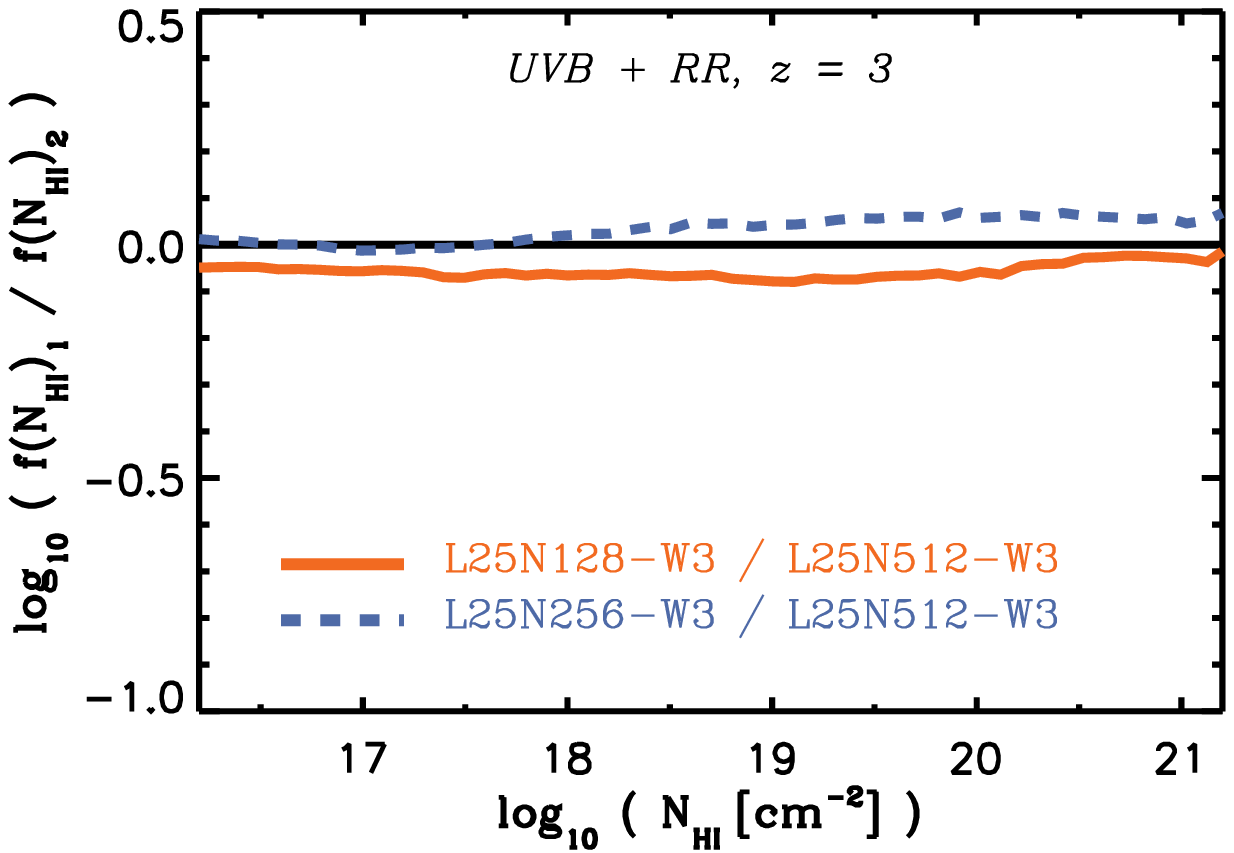}} }
\caption{The relative changes in the HI CDDF using different resolutions, box sizes and cosmologies in the presence of the UVB and diffuse recombination radiation. The \emph{top-left} panel shows the effect of box size on $\fNHI$ for a fixed resolution at $z = 3$, where the orange solid (blue dashed) curve shows the difference between using a box size of $L = 25~(50)$ comoving $\Mpch$ and a box size of $L = 100$ comoving $\Mpch$. The \emph{top-right} panel shows the same effect but for smaller box sizes: the orange solid (blue dashed) curve shows the difference between using a box size of $L = 6~(12)$ comoving $\Mpch$ and a box size of $L = 25$ comoving $\Mpch$. The \emph{bottom-left} shows the effect of using a cosmology consistent with WMAP 3-year results instead of using a cosmology based on the WMAP 7-year constraints. The orange solid and blue dashed curves show this effect for simulations with box sizes of $L = 6$ and $25$ comoving $\Mpch$, respectively. The \emph{bottom-right} panel shows the effect of resolution.}
\label{fig:fNHI-box-res}
\end{figure*}

The size of the simulation box may limit the abundance and the density of the densest systems captured by the simulation. In other words, very massive structures, which may be associated with the highest HI column densities, cannot be formed in a small cosmological box. Indeed, as shown in the top panels of Figure \ref{fig:fNHI-box-res}, one needs to use cosmological boxes larger than $\gtrsim 25$ comoving $\Mpch$ in order to achieve convergence in the $\HI$ distribution (see also \citealp{Altay11}). On the other hand, the bottom-right panel of Figure~\ref{fig:fNHI-box-res} shows that changing the resolution of the cosmological simulations also affects $\fNHI$, although the effect is small.
\par
The adopted cosmological parameters also affect the gas distribution and hence the HI CDDF. For instance, one expects that the number of absorbers at a given density varies with the density parameter $\Ob$, and the root mean square amplitude of density fluctuations $\sigeight$. The bottom-left panel of Figure \ref{fig:fNHI-box-res} shows the ratio of column densities in simulations assuming WMAP 7-year and 3-year parameters. The ratio is only weakly dependent on the box size of the simulation and its resolution. This motivates us to use this ratio to convert the HI CDDF between the two cosmologies for all box sizes and resolutions (at any given redshift). While this is an approximate way of correcting for the difference in the cosmological parameters, it does not affect the main conclusions presented in this work (e.g., the lack of evolution of $\fNHI$).
\section{RT convergence tests}
\label{ap:RT-conv-test}
\begin{figure*}
\centerline{\hbox{\includegraphics[width=0.45\textwidth]
             {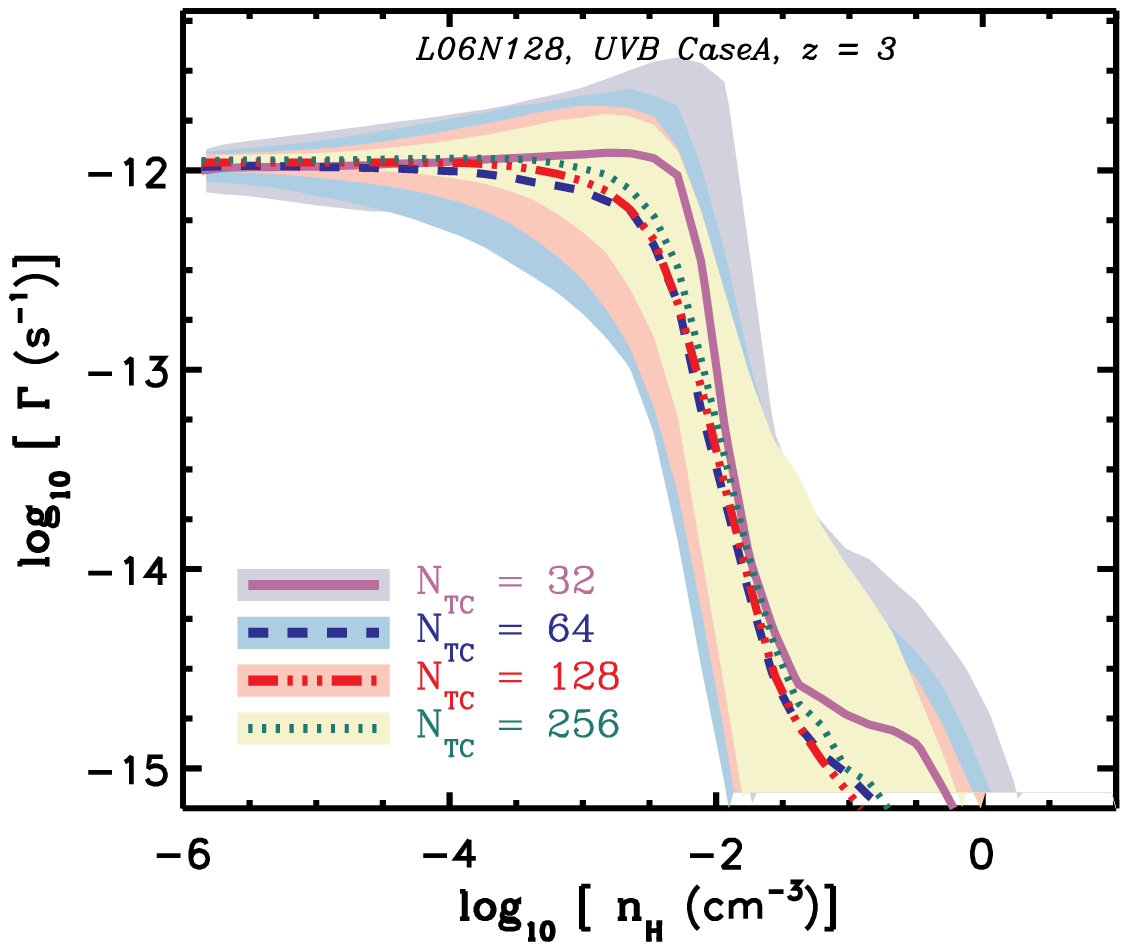}} 
             \hbox{\includegraphics[width=0.45\textwidth]
             {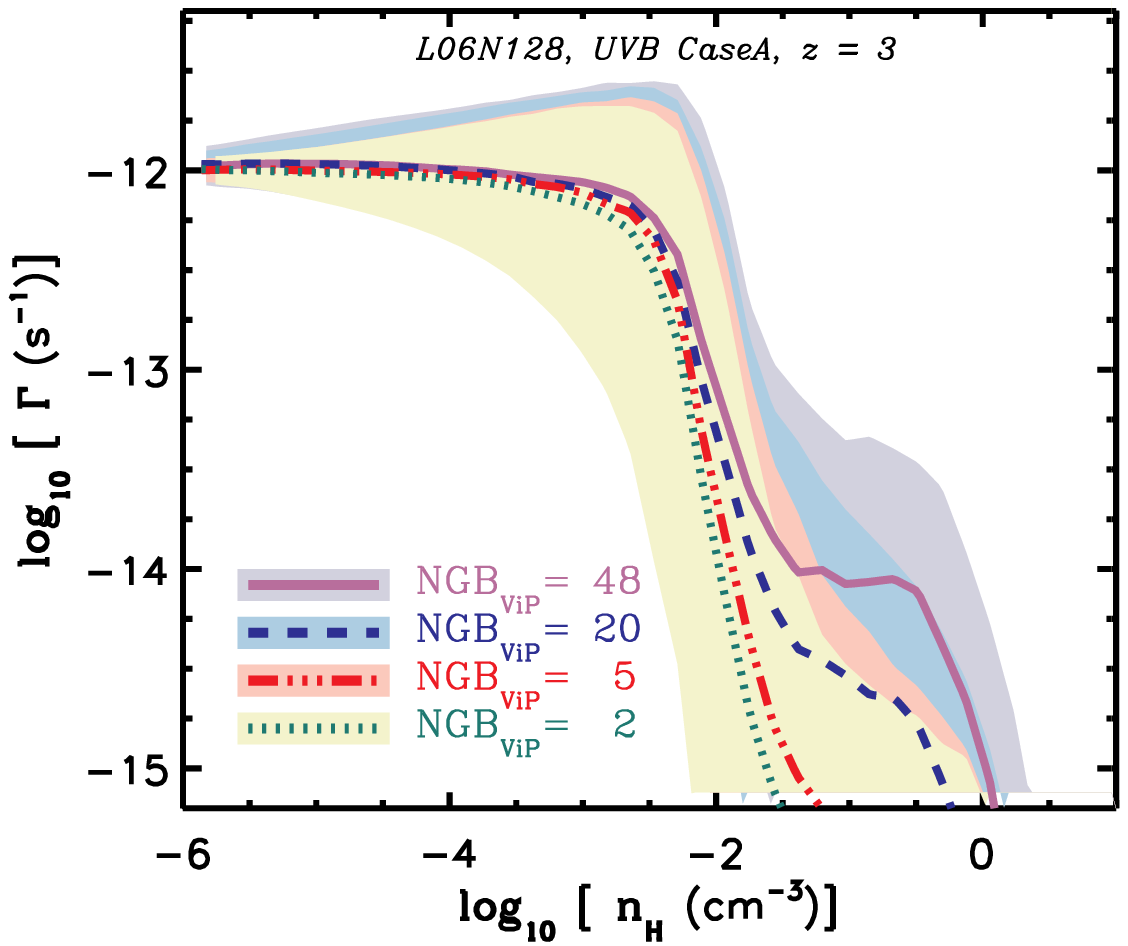}} }
\caption{The UVB photoionization rate is converged for our adopted angular resolution, i.e., $N_{TC} = 64$, as shown in the \emph{left} panel and our adopted number of ViP neighbors, i.e., $\rm{NGB_{\rm{ViP}}} = 5$, as shown in the \emph{right} panel. Photoionization rate profiles are shown for the \emph{L06N128} simulation in the presence of the UVB radiation where the Case A recombination is adopted. The curves show the medians and the shaded areas around them indicate the $15\% - 85\%$ percentiles.}
\label{fig:convtst_UVB}
\end{figure*}
\subsection{Angular resolution}
The left-hand panel of Figure~\ref{fig:convtst_UVB} shows the dependence of
photoionization rates on the adopted angular resolution, i.e., the opening 
angle of the transmission cones $4\pi/N_{\rm TC}$. The photoionization rates are converged
for $N_{\rm TC}=64$ (our fiducial value) or higher.

\subsection{The number of ViP neighbors}
The right panel of Figure~\ref{fig:convtst_UVB} shows the dependence of the photoionization rates on the number of SPH neighbors
of ViPs. As discussed in \S\ref{sec:traphic}, ViPs distribute the ionizing photons they absorb among their $\rm{NGB_{ViP}}$ nearest SPH
neighbors. The larger the number of neighbors, the larger the volume over which photons are distributed, and the more extended is the
transition between highly ionized and self-shielded gas. The photoionization rates converge for $\lesssim 5$ ViP neighbors (our fiducial value is 5). 

\subsection{Direct comparison with another RT method}
\label{ap:Altay-comp}
\begin{figure*}
\centerline{\hbox{\includegraphics[width=0.45\textwidth]
              {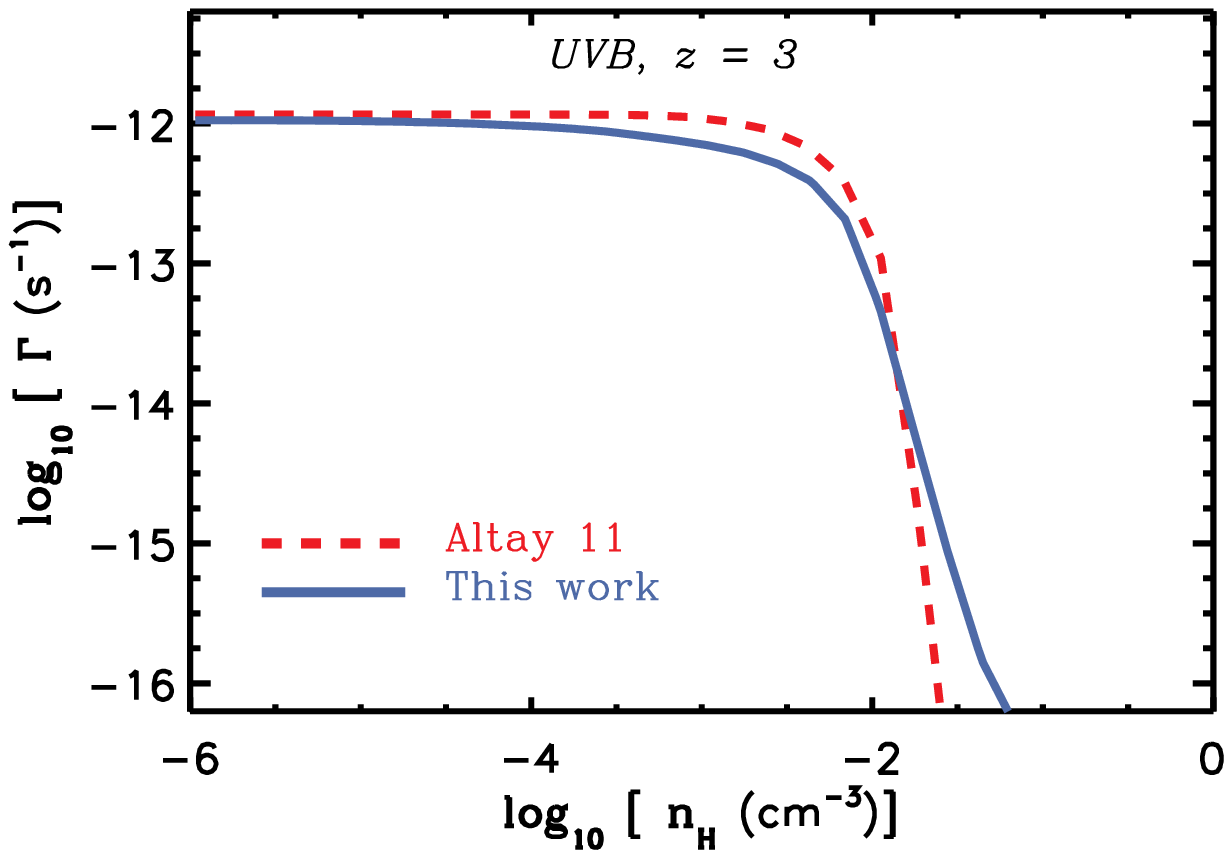}}
             \hbox{\includegraphics[width=0.45\textwidth]
             {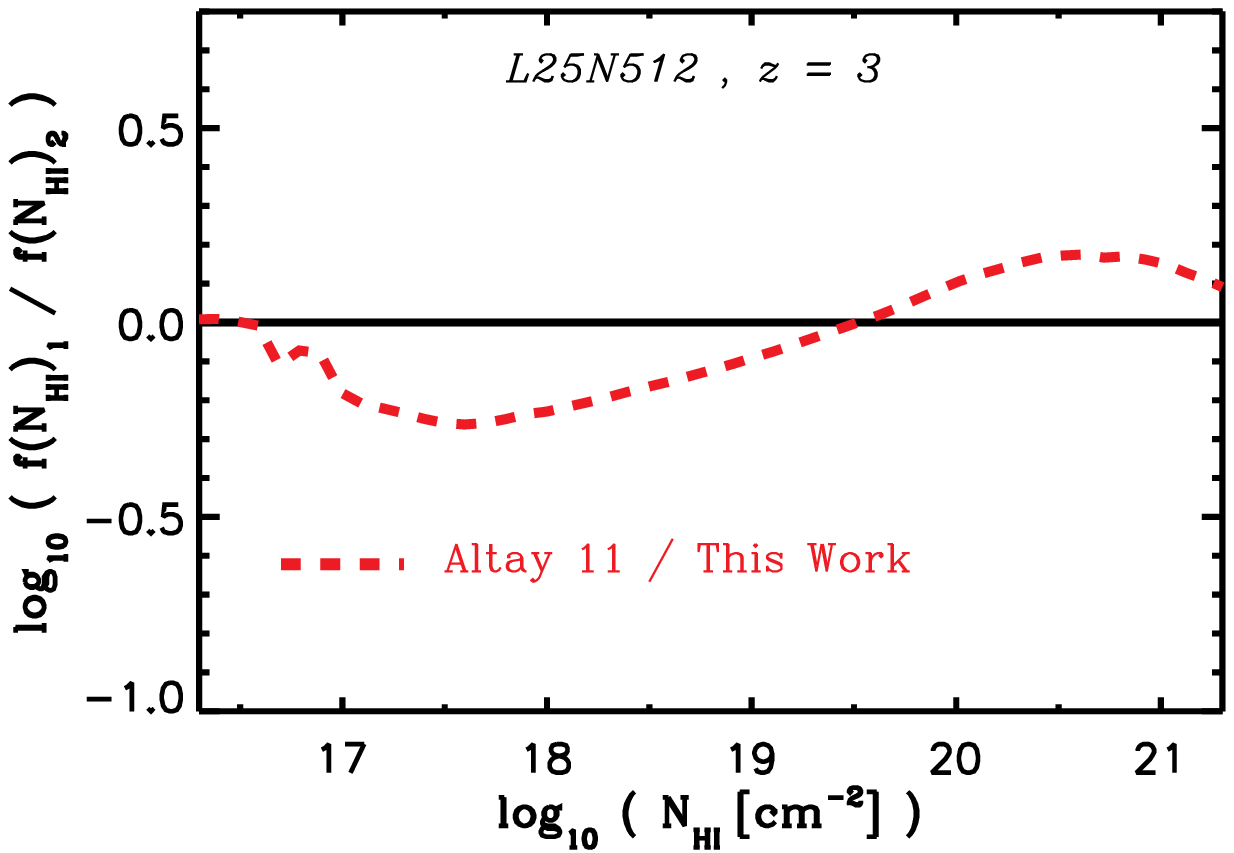}} }
\caption{\emph{Left:} Median UVB photoionization rate as a function of density at $z = 3$ using different RT methods. The red dashed curve shows the results based on the method that has been used in \citet{Altay11} and the blue solid curve shows the result of this work. \emph{Right:} The HI CDDF of the \emph{L25N512} simulation at $z = 3$ using different RT methods and without RR. The red dashed curve shows the ratio between the HI CDDF given in \citet{Altay11} and our results. This comparison shows that despite the overall agreement between our results and \citet{Altay11}, there are some important differences.}
\label{fig:comp_gab}
\end{figure*}

\citet{Altay11} used cosmological simulations from the reference model of the OWLS project \citep{Schaye10}, i.e., a simulation run with the same hydro code as we used in this work, to investigate the effect of the UVB on the HI CDDF at $z = 3$. However, they employed a ray-tracing method very different from the RT method we use here. Furthermore, they did not explicitly treat the transfer of recombination radiation. In Figure \ref{fig:comp_gab}, we compare one of our UVB photoionization rate profiles\footnote{Note that in our simulations the UVB photoionization rate is converged with the box size and the resolution as shown in $\S$\ref{sec:Photoionization-density-fit}.} with the photoionization rate found by \citet{Altay11} in a similar simulation. The overall agreement is very good, but the comparison also reveals important differences.
\par
\citet{Altay11} calculate the average optical depth around every SPH particle within a distance of 100 proper kpc, assuming the UVB is unattenuated at larger distances. Then, they use this optical depth to calculate the attenuation of the UVB photoionization rate for every particle. This procedure may underestimate the small but non-negligible absorption of UVB ionizing photons on large scales. Indeed, by tracing the self-consistent propagation of photons inside the simulation box, we have found that the UVB photoionization rate decreases gradually with increasing density up to the density of self-shielding. However, we note that the small differences between our UVB photoionization rates and those calculated by \citet{Altay11} at densities below the self-shielding, become slightly smaller by increasing the angular resolution in our RT calculations (see the left panel of Figure \ref{fig:convtst_UVB}).
\par
\section{Approximated processes}
\label{ap:caviets}
\begin{figure*}
\centerline{\hbox{\includegraphics[width=0.45\textwidth]
             {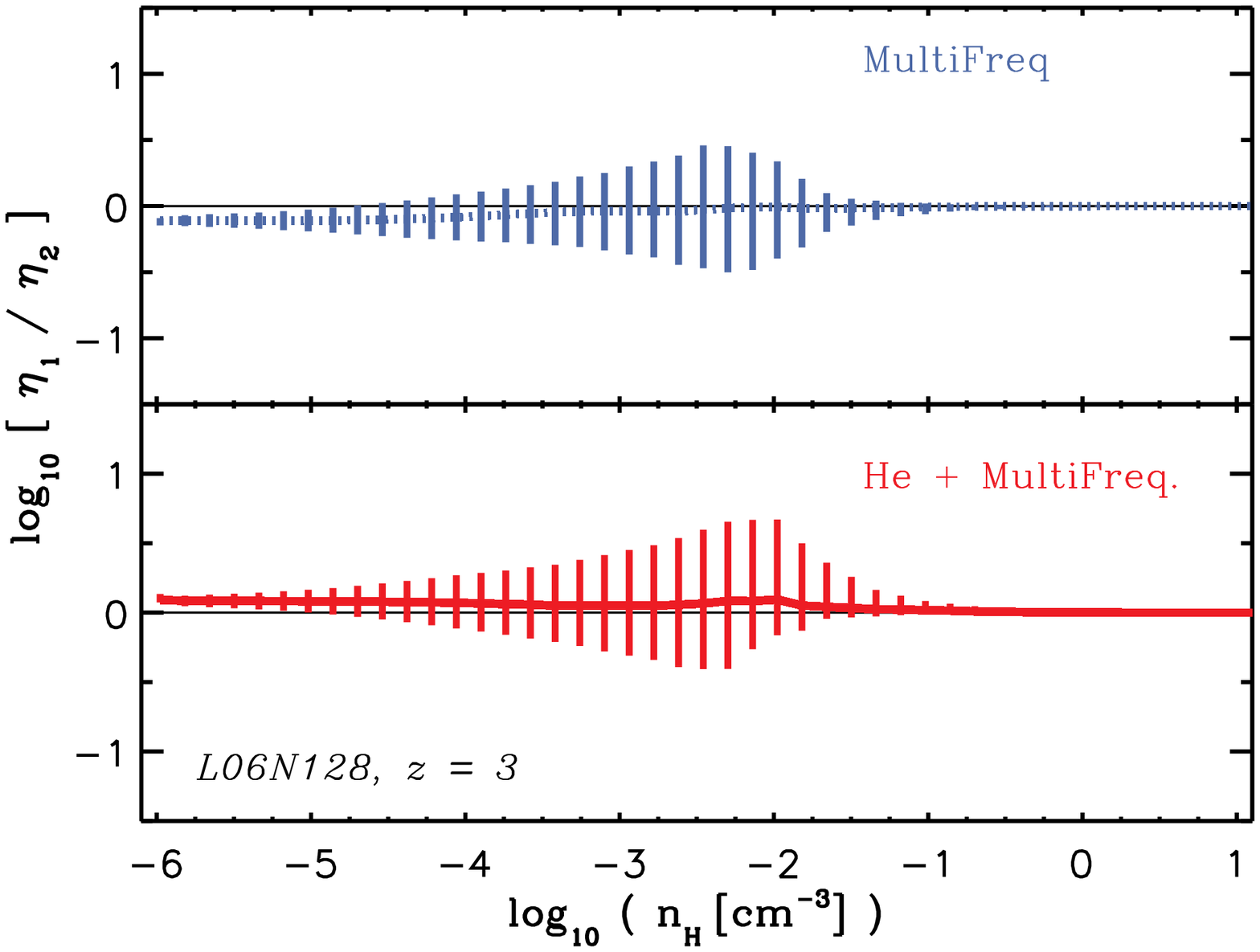}} 
             \hbox{\includegraphics[width=0.45\textwidth]
             {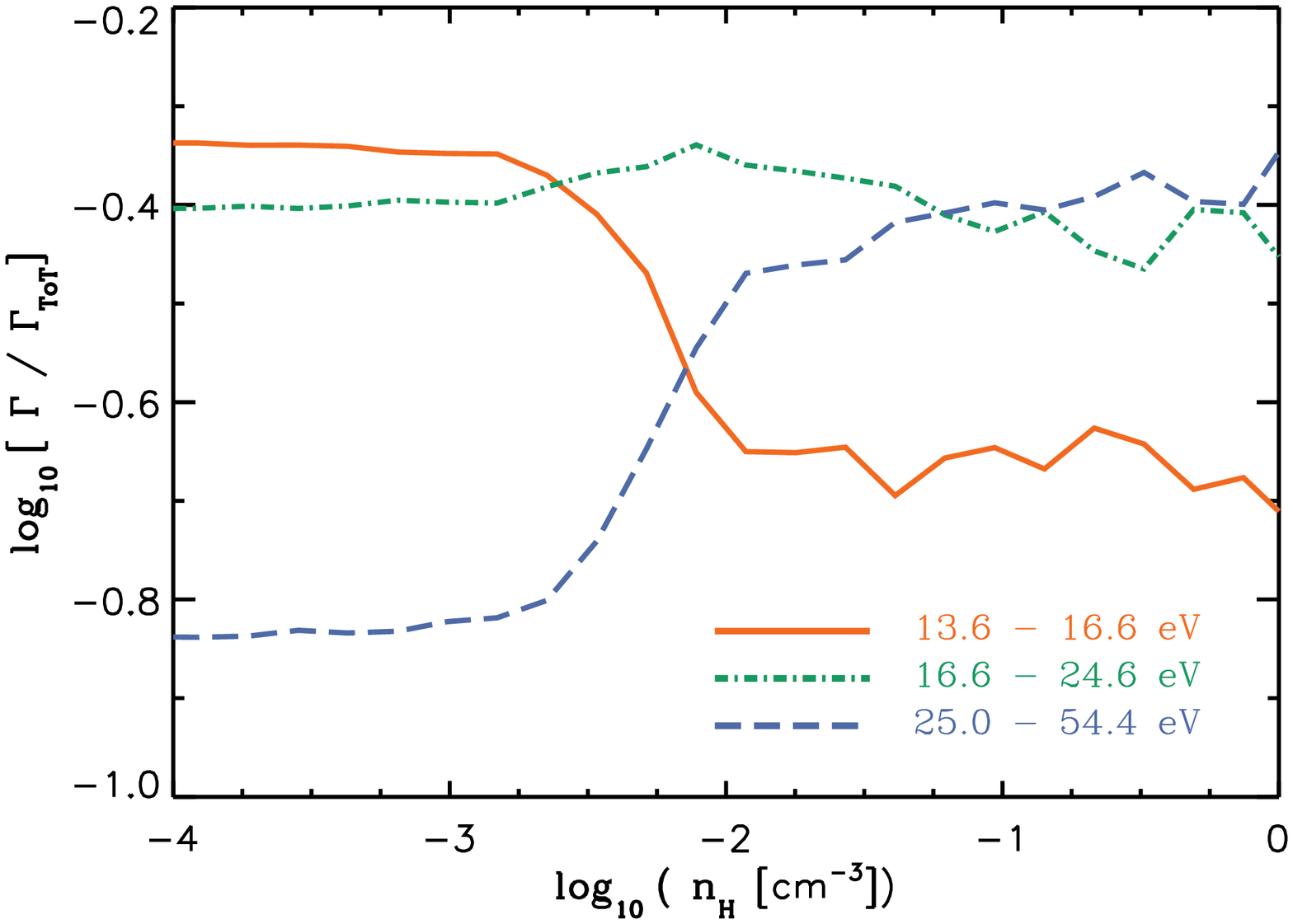}} }
\caption{Spectral hardening and multifrequency treatment do not change the HI distribution significantly. \emph{Left:} The ratio between hydrogen neutral fractions, $\eta$, obtained by using 3 frequency bins and by using the gray approximation is shown in the top-left apnel. The ratio between hydrogen neutral fractions resulting from a simulation with 4 frequency bins and explicit He treatment and the same quantity using the gray approximation and without explicit He treatment is shown in the bottom-left panel. The vertical lines with different lengths indicate the $15\%-85\%$ percentiles. \emph{Right:} The fractional contribution of different frequency bins to the total UVB photoionization rate for the simulations with 3 frequency bins. All the RT calculations are performed using the \emph{L06N128} simulation at $z =3$ in the presence of the UVB and assuming Case A recombination.}
\label{fig:He-Multifreq}
\end{figure*}
\subsection{Multifrequency effects}
As discussed in \S\ref{sec:UVB-normalization}, in our RT simulations we have treated the multifrequency nature of the UVB radiation in the gray approximation (see equation \ref{eq:gray-sigma}). This approach does not capture the spectral hardening which is a consequence of variation of the absorption cross-sections with frequencies. We tested the impact of spectral hardening on the HI fractions by repeating the \emph{L06N128} simulation at $z = 3$ with the UVB using 3 frequency bins. We used energy intervals $\left[ 13.6 - 16.6\right]$, $\left[ 16.6 - 24.6\right]$  and $\left[ 24.6 - 54.4\right]~\rm{eV}$ and assumed that photons with higher frequencies are absorbed by He. The result is illustrated in the top section of the left panel in Figure \ref{fig:He-Multifreq}, by plotting the ratio between the resulting hydrogen neutral fraction, $\eta$, and the same quantity in the original simulation that uses the gray approximation. This comparison shows that the simulation that uses multifrequency predicts hydrogen neutral fractions $< 10\%$ lower at low densities (i.e., $\nH \lesssim 10^{-4}\cmcb$). This does not change the resulting $\fNHI$ noticeably at the column densities of interest here.
\par
The spectral hardening captured in the simulation with 3 frequency bins is illustrated in the right panel of Figure \ref{fig:He-Multifreq}. This figure shows the fractional contribution of different frequencies to the total UVB photoionization rate as a function of density. The red solid curve shows the contribution of the bin with the lowest frequency and drops at the self-shielding density threshold. On the other hand, the fractional contribution of the hardest frequency bin increases at higher densities, as shown with the blue dashed curve. Despite the differences in the fractional contributions to the total UVB photoionization rate, the absolute photoionization rates drop rapidly at densities higher that the self-shielding threshold for all frequency bins. 
\subsection{Helium treatment}
\label{sec:Helium}
A simplifying assumption frequently used in RT simulations which aim to calculate the distribution of neutral hydrogen is to ignore helium in the ionization processes \citep[e.g.,][]{Faucher09, McQuinn10, Altay11}. We adopted the same assumption in our RT calculations which implies that we implicitly assumed the ionization state of neutral helium and its interaction with free electrons to be similar to the trends followed by neutral hydrogen. This has been shown to be a good assumption \citep{Osterbrock06,McQuinn10,Friedrich12}. Nevertheless, we tested the validity of our approximate helium treatment by repeating the \emph{L06N128} simulation at $z = 3$ with the UVB using 4 frequency bins and an explicit He treatment. The first three frequency bins are identical to the bins used in the previous section (i.e., $\left[ 13.6 - 16.6\right]$, $\left[ 16.6 - 24.6\right]$  and $\left[ 24.6 - 54.4\right]~\rm{eV}$) and the last bin is chosen to cover higher frequencies which are capable of HeII ionization. We adopted a helium mass fraction of $25\%$ and a Case A recombination rate. The ratio between the resulting hydrogen neutral fraction and the same quantity when a single frequency is used and helium is not treated explicitly is illustrated in the bottom-left panel of Figure \ref{fig:He-Multifreq}. The hydrogen neutral fractions are very close in the two simulations. However, the simulation with multifrequency and explicit He treatment results in hydrogen neutral fractions that are $< 10\%$ higher at low densities (i.e., $\nH \lesssim 10^{-4}\cmcb$). This difference is barely noticeable in the comparison between the two $\HI$ CDDFs (not shown).
\end{document}